\documentclass[reqno]{amsart}

\usepackage{graphicx}

\newtheorem{lemma}{Lemma}
\newtheorem{theorem}{Theorem}
\newtheorem{corollary}{Corollary}
\newtheorem{hypothesis}{Hypothesis}
\newtheorem{remark}{Remark}
\newtheorem{definition}{Definition}

\newcommand{\be}{\begin{eqnarray}}
\newcommand{\ee}{\end{eqnarray}}
\newcommand{\bee}{\begin{eqnarray*}}
\newcommand{\eee}{\end{eqnarray*}}
\newcommand{\R}{{\mathbb R}}
\newcommand{\N}{{\mathbb N}}

\newcommand{\Ze}{{\mathcal Z}}
\newcommand{\C}{{\mathbb C}}
\newcommand{\I}{{\mathbb I}}
\newcommand{\E}{{\mathcal E}}

\newcommand{\X}{{\mathcal X}}
\newcommand{\Ha}{{\mathcal H}}
\newcommand{\K}{{\mathcal K}}
\newcommand{\uno}{{\mathbb I}}
\newcommand{\resto}{{\mathcal R}}
\newcommand{\U}{{\mathcal U}}
\newcommand{\M}{{\mathcal M}}
\newcommand{\asy}{{\it O}}
\def\norma#1{\left\|#1\right\|}
\newcommand{\im}{{i}}
\newcommand{\e}{{\rm e}}
\newcommand{\di}{{\rm d}}
\newcommand{\calN}{{\mathcal N}}
\newcommand{\B}{{\mathcal B}}

\def\normag#1#2{\left|X_{#1}\right|_{#2}}
\newcommand{\F}{{\mathcal F}}
\newcommand{\G}{{\mathcal G}}
\def\Pc{{\mathcal P}}
\def\Sc{{\mathcal S}}
\def\Tr{{\mathcal T}}
\newcommand{\re}{{\mathbb R}}
\def\XR{\X_{\re}^s}
\def\mus{\left(\frac{\epsilon}{\epsilon_0}\right)}
\def\Ic{{\mathcal I}}

\def\Le{{\mathcal L}}

\begin{document}

\title [Normal forms and NLS]{Exponential times in the 
  one-dimensional Gross--Petaevskii equation with multiple well potential}

\author {Dario Bambusi}
\address {Dipartimento di Matematica\\ Universit\'a degli studi 
di Milano\\ Via Saldini 50, Milano 20133, Italy}

\email {bambusi@mat.unimi.it}

\author {Andrea Sacchetti}

\address {Dipartimento di Matematica Pura ed Applicata\\ Universit\'a 
degli studi di Modena e Reggio Emilia\\ Via Campi 213/B, Modena 41100, Italy}

\email {Sacchetti@unimore.it}

\date {\today}

\thanks {This work is partially supported by the INdAM project {\it
    Mathematical modeling and numerical analysis of quantum systems
    with applications to nanosciences}. \ DB was also supported by
    MIUR under the project COFIN2005 {\it Sistemi dinamici nonlineari
    ed applicazioni fisiche}. \ AS was also supported by MIUR under
    the project COFIN2005 {\it Sistemi dinamici classici, quantistici
    e stocastici}.}

\begin {abstract}
We consider the Gross-Petaevskii equation in 1 space dimension with a
$n$-well trapping potential. \ We prove, in the semiclassical limit,
that the finite dimensional eigenspace associated to the lowest $n$
eigenvalues of the linear operator is slightly deformed by the
nonlinear term into an almost invariant manifold $\M$. \ Precisely, one
has that solutions starting on $\M$, or close to it, will remain close to
$\M$ for times exponentially long with the inverse of the size of the
nonlinearity. \ As heuristically expected the effective equation on
$\M$ is a perturbation of a discrete nonlinear Schr\"odinger equation. \ We 
deduce that when the size of the nonlinearity is large enough then tunneling 
among the wells essentially disappears: that is 
for almost all solutions starting close to $\M$ their restriction  
to each of the wells has norm approximatively constant over the
considered time scale. \ In the particular case of a double well
potential we give a more precise result showing persistence or
destruction of the beating motions over exponentially long times. \ The
proof is based on canonical perturbation theory; surprisingly enough,
due to the Gauge invariance of the system, no non-resonance condition is required.
\end{abstract}


\maketitle

\section {Introduction}

\begin{figure}
\begin{center}
\includegraphics[height=7cm,width=7cm]{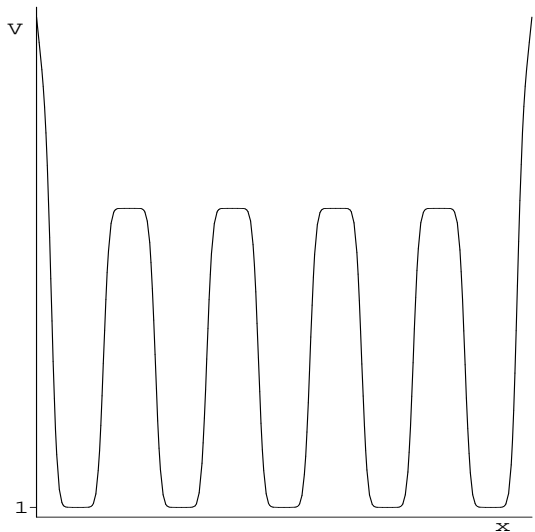}
\caption{Plot of a trapping potential with $n$ wells.}
\label {Fig0}
\end{center}
\end{figure}

In this paper we study the dynamics of low energy states of the one-dimensional 
Gross-Petaevskii equation (hereafter also called nonlinear Schr\"odinger
equation, NLS)
\be
\left \{
\begin {array}{l}
i \hbar \dot \psi^t = H_0 \psi^t + \epsilon \left|\psi^t \right|^{2\sigma}
\psi^t , \ \dot \psi^t = \frac {\partial \psi^t
}{\partial t}, \\ \left. \psi^t (x)\right |_{t=0} = \psi^0 (x) \in L^2
(\R ) , \ \| \psi^0 \|_{L^2} =1 ,
\end {array}
\right. \label {eq1}
\ee
where $\sigma$ is a positive integer number and 
\be 
H_0 = -  {\hbar^2} \frac {d^2 }{d x^2} + V, \ x \in \R , \label {eq2} 
\ee
is the linear Hamiltonian operator and $V(x)$ a $n$--well potential. \ By this we mean that $V$ has $n$ nondegenerate distinct minima $x_1,...,x_n$ where the potential has essentially the
same behavior (e.g. one can assume that its first $r$ derivatives are equal at all the minima, for some positive integer $r\ge 4$). \ We also assume that the potential is trapping, i.e. $V$ tends to infinity as $|x|\to\infty$. \ The situation we have in mind is that of a Bose Einstein condensate trapped by an unbounded potential and also subjected to a periodic-like force field (see Figure \ref {Fig0}); in such a case the parameter $\epsilon$ can be thought as a measure of  the number of particles in the condensate (see, e.g. \cite {TS}). 

Consider first the linearized problem. \ The fundamental state of $H_0$ is approximatively degenerate in the sense that, denoting by 
\bee
\lambda_1<\lambda_2<...<\lambda_n<...
\eee
the eigenvalues of $H_0$, one has 
\bee
\lambda_n-\lambda_1 \ll \lambda_{n+1}-\lambda_n
\eee
in the semiclassical limit, i.e. $\hbar \ll 1$. \ Then the most
interesting situation occurs when the normalized eigenfunctions
$\varphi_1,...,\varphi_n$, corresponding to $\lambda_1,...,\lambda_n$,
are delocalized among the wells. \ Indeed, in
such a case a solution in $\Phi_0:=$span$(\varphi_1,...,\varphi_n) $
performs a quasiperiodic motion and the probability of finding the
particle in any fixed well undergoes great changes over a time scale
of the order of $T=\pi \hbar /\omega $ with $\omega =(\lambda_n -
\lambda_1)/2$. \ In the case of double well potential such a
phenomenon is usually known as beating motion and the beating period
is given by $T:=\pi\hbar/\omega$, with $\omega=(\lambda_2-\lambda_1)/2
$.

The main question is the behavior of the system when the nonlinearity is restored. \ In the case of a double well potential the problem was tackled in a series of papers \cite{AFGST,BaSa,GM,GMS,RSFS,Sa1,Sa2,VA,VA2,Zh}; in particular, it was shown that, up to times of order $T$, the dynamics is well described by an Hamiltonian integrable system with two degrees of freedom obtained by restricting the Hamiltonian, i.e. the energy of the system, to
$\Phi_0$. \ In particular, this result has been used in order to show that the beating motion is generic for values of the nonlinearity strength $\epsilon$ below a certain threshold value, while new localized states appear for larger nonlinearity strength (i.e. as the number of particles of a Bose-Einstein condensate increase) and for even larger values of the
nonlinearity strength the beating motion disappears.

In the case of a multiple well potential the situation was studied e.g. in the paper \cite{TS} where the authors deduced (formally) the discrete nonlinear Schr\"odinger as an effective equation for the dynamics in $\Phi_0$ and used it in order to study some of the features of the
model. We are not aware of rigorous results in the case of a multiple well potential.

In the present paper we study the nonlinear dynamics using the methods of Hamiltonian perturbation theory for PDEs \cite{Bam99,Bam99a,BG04,Kuk} and we prove that the manifold $\Phi_0$, which is invariant for the linear dynamics, is only slightly deformed by the nonlinearity into a new manifold $\M$ which is {\it approximatively invariant} for the complete dynamics. \ By this statement we mean that solutions starting on $\M$, or close to it,
will remain close to $\M$ for times which are exponentially long with $T$; this result improves the one given by \cite {BaSa, Sa1, Sa2} for double well potentials where the control of the approximate solution was given for times of order $T$. \ Moreover we show (see Lemma \ref {discreto}) that the dynamics on $\M$ is described at first order by a discrete nonlinear Schr\"odinger equation
\be
\label{DNLSi}
i\dot \psi_j=\delta_j\psi_j+\Lambda_j\psi_{j+1}+\Lambda_{j-1}\psi_{j-1}+
\eta\psi_j\left|\psi_j\right|^{2\sigma}\ ,\ j=1,...,n ,
\ee
where $\psi_0=\psi_{n+1}:=0$, $\Lambda_j,\delta_j,\eta$ are suitable constants and where, having in mind the case of Bose-Einstein condensates, $|\psi_j|^2$ represents the fraction of
particles in the $j$-th well for $j=1,...,n$ (see Sect. \ref{s.DNLS} for their precise definition). \ In particular, it is quite easy to study the discrete NLS \eqref{DNLSi} from
the anticontinuum limit \cite{MA} obtaining that when $\eta$ is large enough then solutions corresponding to almost all initial data have the property that $|\psi_j|^2$ is essentially a constant of motion. \ In the case of double well potential, we get that the dynamics on $\M$ is
(up to an exponentially small error) that of an integrable Hamiltonian system with two degrees of freedom. \ This allows us to control also the trajectories of the solutions on $\M$ showing that for small $\eta$ the beating phenomenon persists for exponentially long times, while
for higher values of $\eta$ only motions which are essentially localized in one of the two wells exist, at least for exponentially long times.

We emphasize that from the technical point of view the main result
(Theorem \ref {Theorem2}) is quite surprising since the application of
canonical perturbation theory is typically possible only when some
non-resonance conditions are satisfied. \ On the contrary, here the
result is valid for {\it any} multiple well potential, whose
eigenvalues might fulfill arbitrary resonance
conditions (i.e. the eigenvalues can be linearly dependent
over the relative integers). \ This is possible since NLS is an
infinite dimensional Gauge invariant Hamiltonian system. \ To explain
how this property is exploited we recall that canonical perturbation
theory allows us to remove from the Hamiltonian all non-resonant
monomials. \ In particular, given an arbitrary monomial it can be
eliminated if it is non-resonant. \ However, in NLS only Gauge
invariant monomials appear, and we will show that, for a Gauge
invariant monomial, the non-resonance condition is (almost) trivially
fulfilled. \ Actually, in order to avoid any restriction on the
potential we have to use a resonant construction which is not self
evident a priori (see Subsection \ref{NonCoupling}). \ We think that
these ideas could be useful in the study of Hamiltonian systems with
symmetry, and possibly also for the investigation of further dynamical
properties of NLS. \ A further technical ingredient which is
fundamental for the proof is the use of Sobolev like spaces
constructed as the domains of the powers of $H_0$. \ To use such
spaces one has to show that they form Banach algebras under the
pointwise multiplication. \ Here we give a detailed proof of such a
property that we think could be useful in further investigations of
NLS.

The paper is organized as follows. \ In Section \ref{Mresult} we state
our main results (Theorem \ref {Theorem2}, Theorem \ref {Theorem3} and
their corollaries). \ Section \ref{Mresult} is
divided into 5 subsections: in Subsection \ref{linear} we review some
known facts about the structure of the low lying eigenvalues of the
linear Schr\"odinger equation with an $n$ well potential; in Subsection
\ref{posedness} we introduce the Sobolev like spaces in which NLS
equation will be studied and give their main properties; in Subsection
\ref{esp} we will state the result on the approximate invariant
manifold; in Subsection \ref{s.DNLS} we will give the effective
equation on the approximatively invariant manifold and deduce the
localization properties of the solutions; finally in Subsection
\ref{s.double} we will study the particular case of a double well
potential.\ In Section \ref{s.proof} we prove our main results. \ This
section is also divided into 6 subsections that correspond to the
different parts of the proof. \ The proof of Theorem 1 (algebra
property of Sobolev like spaces) and of some technical Lemmas are left
to appendixes A and B respectively.

{\it Acknowledgments.} DB would like to thank Panos Kevrekidis for
pointing to his attention the paper \cite{TS} and the connection
between the Gross-Pitaevskii and the discrete NLS equations.

\section {Main results}
\label{Mresult}

\subsection{Linear theory}\label {linear}

\begin {hypothesis} \label {Hypothesis2}

The potential $V (x)\in C^\infty (\R )$ is a real valued function such that:

\begin {itemize}

\item [i.] $V (x)$ admits $n$ minima at $x_1<x_2<...<x_n $ such that 
\be 
V (x) > V_{min} = V(x_j )=1 , \ \ \forall x \in \R , \ x\not= x_j ,\ 
j=1,...,n; \label {eq33} 
\ee
\item [ii.] There exists a constant $C>1$ such that  
\bee
C^{-1} \langle x \rangle^2 \le V(x), \ \langle x \rangle = \sqrt {1+x^2}; 
\eee
\item [iii.] There exists a positive $m\geq2$ such that for any  $k\in \N$ %
\bee 
\left | \frac {d^{k} V(x)}{dx^k} \right | \le C_k \langle x \rangle^{m-k} 
\eee
for some positive constant $C_k$;
\item [iv.] The minima are nondegenerate and 
\be
\label{eq35}
\frac {d^2 V(x_j)}{dx^2}=C >0 
\ee
with $C$ independent of $j$;
\item [v.] The shape of the potential at the bottom of the minimum $x_j$ is approximatively independent of $j$; precisely: there exists $r \geq 4$ such that
\be
\frac{d^kV}{dx^k}(x_j)=\frac{d^kV}{dx^k}(x_i),\ \ \forall i,j =1,2,\ldots , n , \ \ \forall k=2,...,r. \label {eq36}
\ee
\end {itemize}

\end {hypothesis}

Hereafter, $C$ will always denote a positive, typically large, constant whose value changes from line to line, and which is independent of $\hbar$, $\epsilon$ and $t$.

The operator $H_0$ formally defined by (\ref {eq2}) admits a self-adjoint realization (still denoted by $H_0$) on $L^2 (\R )$ (Theorem III.1.1 in \cite {BS}) with purely discrete spectrum. \ Let $\lambda_k$, $k \in \N$, be the non degenerate eigenvalues of $H_0$
\bee
\lambda_1 < \lambda_2 < \lambda_3 < \lambda_4 < \ldots < \lambda_k < \ldots , \ \lambda_k    \stackrel{k \uparrow \infty}{\rightarrow}  \infty, 
\eee
with associated normalized (in $L^2$) eigenvectors $\varphi_k (x)$; the set $\{\varphi_k (x) \}_{k=1}^\infty$ is an orthonormal base of $L^2$. 

The lowest part of the spectrum can be studied in the semiclassical limit using the construction of \cite{He}, that we shortly recall. 

Having fixed a positive constant $a>1=V_{min}$ we consider the set $V^{-1}((-\infty,a))$ and we assume that $a$ is such that this set is the union of $n$ disjoint open sets $\U_j$ with $x_j\in\U_j$. \ Having fixed $j$ we consider the operator formally defined on $L^2$ as
\bee
H_j = - \hbar^2 \frac {d^2}{dx^2} + V_j 
\eee
where $V_j$ is a modified potential defined as 
\be
\label{eq41}
V_j(x)=
\left \{ 
\begin{array}{ll}
V(x)& \text{ for } \  x\in\U_j \\
\max [ a,V(x)] & \text{ for }\ x\not \in\U_j
\end{array}   
\right.
\ee
Let $\hat \lambda_{j} $ be the lowest eigenvalue of $H_j$ with
associated normalized eigenvector $\hat \varphi_j$. \ Using the
semiclassical construction of the eigenvalues close to the bottom of a
well (see e.g. \cite{BGP,Sj}) one has that assumption Hyp. \ref
{Hypothesis2} v., implies that
\bee
\hat \lambda_j = 1 + O(\hbar ) , \ |\hat \lambda_{j} -\hat \lambda_{i} | \le C \hbar^{r/2} 
\eee 
and 
\bee
\| \hat \varphi_j (x) - \hat \varphi_i (x+x_i-x_j)\|_{L^2} \le C
\hbar^{r/2},  
\eee
for any $ i,j=1,2,\ldots ,n$. 

Let 
\bee
\Gamma_j = \int_{x_j}^{x_{j+1}} \sqrt {V(x)-1} dx , \ j=1,2,\ldots , n-1, 
\eee
be the Agmon distance among the two minima $x_j$ and $x_{j+1}$. \
Let $\Gamma $ any fixed positive real number such that $\Gamma <
\min_j \Gamma_j$. Then, from the theory of \cite{He} there exist
some constants $c_{ij}$ such that for any $i$ and $j$
\be
&& \norma{\varphi_j-\sum_{k=1}^nc_{kj}\hat\varphi_k}_{L^2} \leq Ce^{-\Gamma/\hbar}, \label{agg.1} \\ 
&& \norma{\hat\varphi_j\hat\varphi_i}_{L^\infty} \leq Ce^{-\Gamma/\hbar}, \ i\not=j , \label {agg.1bis} \\
&& |\lambda_{i_j} - \hat \lambda_{j} | \le C e^{-\Gamma/\hbar}\ \text{ for some }\
i_j\in\left\{1,...,n\right\} .\label{agg.2}
\ee
Moreover, the projector $\hat \Pi$ on $\hat\Phi_0:= $span($\hat
\varphi_j$) is a bijection among $\Phi_0:= $span($\varphi_j$) and
$\hat\Phi_0$ itself. \ Remark also that one can choose the functions
$\hat\varphi_j$ to be real valued.

\ Hence, the lowest $n$ eigenvalues of $H_0$
fulfill
\bee 
C^{-1} \hbar < \lambda_{j} -1 < C \hbar\ ,\quad j=1,...,n ,
\eee
for some $C>1$ and 
\be 
\omega = \frac 12 (\lambda_n-\lambda_1)\leq C\hbar^{r/2} \label {eq7} 
\ee
Furthermore, making use of the same arguments, it follows also that
\be
\inf_{j=1,\ldots , n}\ \inf_{\lambda \in \sigma (H_0) - \{ \lambda_1,...,\lambda_n \} }
[\lambda - \lambda_{j} ] \ge C^{-1} \hbar  \label {eq6} 
\ee
We also use the following projectors
\bee \Pi =\sum_{j=1}^n \langle \varphi_j , \cdot \rangle \varphi_j \ \
\mbox { and } \ \ \Pi_c = {\mathbb I} - \Pi \ \ \mbox { and } \ \ \Phi_0=\Pi L^2 ; 
\eee

\begin{remark}
\label{agg.3}
As already emphasized in the introduction the most interesting situation occurs when the true eigenfunctions $\varphi_j$ are delocalized between the wells, a property that occurs if the potential is exactly periodic in some region or more generally if the order of magnitude of some the off diagonal elements of the matrix formed by the constants $c_{ij}$ is of the same order of
magnitude as the difference between the approximate eigenvalues $\hat \lambda_{j}$, a property that is quite difficult to ensure in a general situation. \ For this reason, essentially in order to fix ideas, we decided to state our assumptions in the form (i-v) above. 
\end{remark}

\subsection{The nonlinear system: Analytic framework and well posedness.} 
\label{posedness}

\begin{definition} 
For any integer $s \ge 0$ define the Hilbert space $\X^s:=D(H_0^{s/2})$ endowed by the graph norm; more precisely in $\X^s$ we will use the following norm equivalent to the graph norm
\be
\label{eq12}
\norma{\phi}_s^2:=\norma{H_0^{s/2}\phi}^2_{L^2}\equiv \left\langle H_0^s\phi,\phi \right\rangle_{L^2}=\int_{\R}\bar\phi H_0^s\phi \di x, \ \phi \in \X^s .
\ee
\end{definition}

The main step for the proof that the spaces $\X^s$ form a Banach algebra under the pointwise multiplication is the following Theorem.

\begin{theorem}
\label{Theorem1}
Let $s$ be any positive integer number. \ For small enough $\hbar$ the two norms
\be 
\norma{\phi} ^2_s \ \ \mbox { and } \ \ \| (-\hbar^2\Delta)^{s/2} \phi \|^2_{L^2} +\| V^{s/2} \phi \|^2_{L^2} \label{eq13} 
\ee
are equivalent with an $\hbar$ independent constant.
\end{theorem}

The proof, which is a semiclassical variant of the proof of Lemma 7.2 of \cite{YZ04}, is deferred to Appendix \ref {appendiceA}.  

\begin{remark}
\label{Remark2}
In particular one has that a function $\phi$ is in $\X^s$ if and only if it belongs to the Sobolev space $H^s$ and it decays at infinity so fast that $|\phi|^2 V^s$ is integrable.
\end{remark}

In the spaces $\X^s$ with $s\geq 1$ the system \eqref{eq1} is semilinear, since, using \eqref{eq13} and Gagliardo-Niremberg inequality one has

\begin{corollary} \label{Corollary1}
For any integer $s\geq 1$ there exists a positive constant $C_s$ independent of $\hbar$, such that
\be \norma{\phi_1\phi_2} _s \le C_s \hbar^{-1/2} \norma{\phi_1}_s
\norma{\phi_2}_s ,\ \ \forall \phi_1,\phi_2 \in\X^s ,\ \hbar\ll 1 .\label
{eq14} \ee
Moreover, the map
\bee \X^s\times\X^{s}\ni(\phi,\bar\phi) \mapsto
|\phi|^{2\sigma}\phi\in\X^s \eee
is entire analytic map for any $\hbar>0$ and fulfills
\be
\label{eq15} \norma{ |\phi|^{2\sigma}\phi }_s\leq {C_s}{\hbar^{-\sigma }}\norma{\phi}_s^{2\sigma+1} .
\ee
\end{corollary}

\begin{remark}
\label{Remark3}
In dimension $d>1$  this result remains valid provided $s> d/2$.
\end{remark}

Then by standard Segal theory (see e.g. \cite{Pazy}) the system \eqref
{eq1} is locally well posed in all the spaces $\X^s$ with $s\geq 1$. \
Actually $s>d/2$ is enough, but our proof only applies to integer
values of $s$; in \cite{YZ04} a Strichartz inequality argument was
used to show that it is also locally well posed (LWP) in $\X^s$ with
some $s$ smaller than $d/2$.

From now on we assume that the index $s$ of the space is a fixed
positive integer number and fulfills the condition $s\geq 1$. \ In the
following, in order to fix ideas, one can just think of the case
$s=1$. \ We will denote by $d(.;.)$ the distance in the norm of
$\X^s$.

\subsection{The nonlinear system: Approximatively invariant manifold.}
\label{esp}

In order to state our main result we assume that the size of the nonlinearity is small enough, i.e. $|\epsilon | \ll \hbar^\sigma $, and we introduce the small parameter.
\be
\label{eq19}
\mu:=\omega+\frac{|\epsilon|}{\hbar^\sigma} \ll 1 ,
\ee
where $\omega$ was defined in \eqref{eq7}.

\begin{theorem}
\label{Theorem2}
Consider the system \eqref{eq1} and fix a positive $s\geq 1$. There
exists a positive $\mu_*$ such that, if $\mu<\mu_*\hbar^{3/2}$,
then there exists
a manifold $\M$ (dependent on all the parameters of the system) with
the following properties:

\begin {itemize}

\item [i.]  $\M$ is close to $\Phi_0$, i.e.
\be
\label{eq20}
d(\Phi_0,\M)\leq C\frac{\mu}{\hbar^{3/2}} ,
\ee
where 
\bee d (\Phi_0,\M ) = \sup_{\psi \in \Phi_0}\inf_{ \varphi \in \M} \|
\psi - \varphi \|_s, \eee
and where $\| \varphi \|_s $ is the norm \eqref{eq12}.

\item [ii.] Let 
\bee
d_0 = d (\psi^0 , \M ) = \inf_{ \varphi \in \M} \| \psi^0 - \varphi \|_s 
\eee
be the initial distance from $\M$ and let 
\be
\label{eq21}
\delta = \max \left \{ d_0, \exp \left [ - \frac{\mu_*\hbar^{3/2}}{2\mu}\right ] \right \}
\ee
Then for all times $t$ fulfilling
\be
\label{eq22}
|t|\leq\frac{1}{C\hbar\mu \delta}
\ee
one has 
\be
\label{eq23}
&&d(\psi^t ,\M)\leq C\delta
\\
\label{eq24}
&&\norma{\Pi_c\psi^t}_s\leq C\frac{\mu}{\hbar^{3/2}}
\ee

\end {itemize}

Such a manifold $\M$ is called an \emph {approximatively invariant manifold}.
\end{theorem}

\begin{remark}
\label{Remark4}
The most interesting cases are when 
\bee \delta =\exp \left [ -\frac{\mu_*\hbar^{3/2}}{2\mu} \right ] \ \
\mbox { or } \ \ \delta=C\frac{\mu}{\hbar^{3/2}} .  \eee
Indeed in the first case the time of validity of all the estimates is
exponentially long, while in the second case it is easy to obtain the
following Corollary.
\end{remark}

\begin{corollary}
\label{Corollary2}
Assume that $\norma{\Pi_c\psi^0}_s\leq C\mu\hbar^{3/2}$ then, up to
the times
\be
\label {eq26}
|t|\leq\frac{\hbar^{3/2}}{C\mu^2}
\ee
the estimate \eqref{eq24} holds. 
\end{corollary}

\begin{remark}
\label{Remark5}
This Corollary is a direct extension of the results of
\cite{GMS,Sa1,Sa2} in which the estimate \eqref{eq24} has been proved
(for the double well potential) for a time scale of order
\bee T = \frac {\pi \hbar}{\omega} \le C \frac {\hbar}{\mu} \ll
\mu^{-2}\hbar^{3/2}.  \eee
The improvment is due to the fact that our construction implies that
$\M$ is linearly stable up to an exponentially small error.
\end{remark}

\subsection{The nonlinear system: discrete NLS and suppression of tunneling}
\label{s.DNLS}

To start with we remark that NLS is a Hamiltonian system (see Subsection \ref{hamilton} for a precise description) with Hamiltonian function given by
\be
\label{eq16}
\E(\psi,\bar\psi) := \E_0(\psi,\bar\psi) + \epsilon {\mathcal P}_0 (\psi , \bar \psi )
\ee
where
\bee 
\E_0(\psi,\bar\psi) := \int_{\R}\bar \psi(x)H_0\psi(x) dx
\eee
and
\bee {\mathcal P}_0 (\psi , \bar \psi ) := \frac{1}{\sigma+1} \| \psi^{\sigma+1} \|_{L^2}^2:= \frac{1}{\sigma+1} \int_{\R}\bar \psi^{\sigma+1}\psi^{\sigma+1} dx 
\eee
The general idea is that for initial data close to $\Phi_0$ the system
should be well described by the Hamiltonian system obtained by
restricting $\E$ to $\hat \Phi_0$ which is close to $\Phi_0$. Denote
\bee
\psi =\sum_{j=1}^{n}\psi_j\hat\varphi_j \in \hat \Phi_0 
\eee
then we have the following

\begin {lemma} \label {discreto}  The restriction of \eqref{eq16} to
  $\hat\Phi_0$ takes the form  
\bee \left. \E \right |_{\hat\Phi_0} &=&
\sum_{j=1}^{n}(\Omega+\nu_j)|\psi_j|^2 +\epsilon c\sum_{j=1}^{n}
|\psi_j|^{2\sigma+2} +\sum_{j=2}^{n} c_j\left [  \bar \psi_j\psi_{j-1}
+ \psi_j \bar\psi_{j-1} \right ] + \\ && \ \
+\asy(e^{-2\Gamma/\hbar}) + \epsilon \asy(e^{-\Gamma/\hbar}) \eee
where $\Gamma$ was introduced before equation \eqref{agg.1} and 
\bee
\Omega = \frac 1n \sum_{j=1}^{n}\lambda_j\ ,\ \nu_j:=\int_{\R}\bar{\hat\varphi}_jH_0
\hat\varphi_j dx-\Omega=\asy(\hbar^{r/2})  
\eee
and 
\bee
c=c(\hbar):= \frac{1}{\sigma +1}\frac1n \sum_{j=1}^{n}\| \hat
\varphi_j\|^{2\sigma +2}_{L^{2\sigma +2}} =\asy \left ( \hbar^{-\sigma
/2} \right )
\eee 
and
\bee 
c_j:=\int_{\R}\bar{\hat\varphi}_{j}H_0 \hat\varphi_{j-1} dx =\asy(e^{-\Gamma/\hbar})
\eee
\end {lemma}

\proof Indeed, we have that
\bee
\left. \E \right |_{\hat\Phi_0} = \left. \E_0 \right |_{\hat\Phi_0} +\epsilon \left. {\mathcal P}_0  \right |_{\hat\Phi_0}.
\eee
The first term takes the form 
\bee
\left. \E_0 \right |_{\hat\Phi_0} = \sum_{j=1}^n |\psi_j|^2 d_{jj} + \sum_{|i-j|=1} \bar \psi_i \psi_j d_{ij} + \sum_{|i-j|\ge 1} \bar \psi_i \psi_j d_{ij}
\eee
with
\bee
d_{ij} =\int_{\R}\bar{\hat\varphi}_i H_0 \hat\varphi_j dx  
\eee 
and where (see, e.g., \cite {He,Sa2}) $d_{i,j} =
\asy(e^{-\Gamma/\hbar})$ when $|i-j|=1$ and $d_{i,j}=
\asy(e^{-2\Gamma/\hbar})$ when $|i-j|> 1$. \ For what concerns the
second term, following \cite {Sa2} and making use of \eqref
{agg.1bis}, we have that
\bee
\left. {\mathcal P}_0  \right |_{\hat\Phi_0} = \frac {1}{\sigma +1} \sum_{j=1}^n \| \hat \varphi_j \|_{L^{2\sigma +2}}^{2\sigma +2} |\psi_j |^{2\sigma +2} + \asy(e^{-\Gamma/\hbar})
\eee
where $\| \hat \varphi_j \|_{L^{2\sigma +2}}^{2\sigma +2}= \asy \left ( \hbar^{-\sigma
/2} \right )$ is approximatively independent of $j$. Remark also that,
since the functions $\hat \varphi_j$ are real valued, the quantities
$c_j$ turn out to be real.\qed

\vskip20pt
Hence, up to higher order terms, to the Gauge transformation $ \psi \to e^{i\Omega t/\hbar} \psi $ and to a rescaling of time $t\to \omega t/\hbar$ the restriction of \eqref{eq16} to $\hat \Phi_0$ is given by
\be
\label{eq17}
\K_0:= \sum_{j=1}^{n}\delta_j|\psi_j|^2 +\eta \sum_{j=1}^{n}
|\psi_j|^{2\sigma+2} +\sum_{j=1}^{n+1}\Lambda_j(\bar \psi_j\psi_{j-1}
+\psi_j\bar\psi_{j-1}) , 
\ee
where
\bee 
\eta = \frac {\epsilon c}{\omega}=\asy\left(\frac{|\epsilon|}{\omega \hbar^{\sigma/2}}\right)\ ,  \ \delta_j:=\frac{\nu_j}{\omega}=\asy(1)\ ,\ \Lambda_j:=\frac{c_j }{\omega}=\asy \left( \frac{e^{-\Gamma/\hbar}}{\omega}\right)
\eee
and the equation of motion of \eqref{eq17} are given by 
\bee
\dot \psi_j=-{i} \frac{\partial \K_0}{\partial\bar \psi_j}, \ \mbox { with } \ j=1,...,n. 
\eee
The system \eqref{eq17} has an integral of motion (the restriction of the square
of the $L^2$ norm to $\hat \Phi_0$) given by
\be
\label{eq18}
\Ic:=\sum_{j=1}^{n}|\psi_j|^2
\ee

We analyze now the consequences of our Theorem for the dynamics. \
From the proof of Theorem \ref {Theorem2} we will be able to exactly
describe the restriction of the system to $\M$. \ Actually, we can
state the following Theorem \ref{Theorem3}, which is a result of
Theorem \ref{Theorem6} stated below.

\begin{theorem}
\label{Theorem3}
Under the same assumptions of Theorem \ref{Theorem2} there exists an
analytic canonical transformation $\Tr:\U\to\X^s$, with $\U$ an open
neighborhood of $\Phi_0$ with size independent of $\mu$ and
$\hbar$, such that
\bee
\E\circ\Tr=\Omega\Ic+\omega \K+\tilde\resto  
\eee
where

\begin{itemize}

\item [i.] $\K\circ\hat\Pi=\K$, i.e. $\K$ depends only on the variables $\psi_1,...,\psi_n$,

\item[ii.] $\left\{\Ic,\K \right\}\equiv0$ i.e. $\Ic$ is an integral of motion for the system with Hamiltonian $\K$

\item [iii.] $\K=\K_0+\asy (\mu/\hbar^{3/2})$

\item [iv.] $\tilde\resto=\asy \left [ \exp-\left(\frac{\mu_*\hbar^{3/2}}{2\mu}\right)\right
]+\asy\left(\norma{\Pi_c\psi}_s^2\right)$ and similar estimates hold for its vector field. 

\item[v.] $ \Tr=\I+\asy(\mu/\hbar^{3/2}) $ i.e. the transformation is close to identity.

\end{itemize}
\end{theorem}

\begin{remark}
\label{app.9}
In this framework the manifold $\M$ turns out to simply be $\M=\Tr(\Phi_0)$. 
\end{remark}

In particular it follows that the dynamics on $\M$ is, up to a small
error, the same of a Hamiltonian system with Hamiltonian function
close to $\K_0$ with an integral of motion given by $\Ic$. \
Thus, it is possible to deduce the following Corollary of Theorem
\ref{Theorem7} below, which is particularly relevant in the double
well case.

\begin{corollary}
\label{Corollary3}
Under the same assumptions of Theorem \ref{Theorem2} one also has
\begin{equation}
\label{agg.23}
\left|\Ic(t)-\Ic(0)\right|\leq C\frac{\mu}{\hbar^{3/2}}\ ,\quad
\left|\K_0(t)-\K_0(0)\right|\leq C\frac{\mu}{\hbar^{3/2}}
\end{equation}
up to the times \eqref{eq22}.
\end{corollary}

We focus now on the exponentially long time scale, thus we assume that the quantity $\delta$ of Theorem \ref{Theorem2} is given by 
\bee
\delta =\exp \left [ -\frac{\mu_*\hbar^{3/2}}{2\mu} \right ] 
\eee
and from \eqref{eq22} it follows that up to times of order $e^{\frac {\mu_*\hbar^{3/2}}{2\mu} } $ the vector field $X_{\tilde\resto}$ of $\tilde\resto $ fulfills the a priori estimate
\bee
\norma{X_{\tilde\resto}}_s\leq C \exp \left [ -\frac{\mu_*\hbar^{3/2}}{\mu} \right ]
\eee
Thus, up to an exponentially small drift, a Gauge transformation and a rescaling of time the dynamics on $\M\equiv\Tr(\Phi_0)$ is that of the $n$--dimensional Hamiltonian system $\K$, which is a perturbation of $\K_0$, i.e. of the discrete NLS \eqref {eq17}. 

It is worth mentioning that when $\eta=0$ the dynamics of $\K_0$ is
that of $n$ decoupled harmonic oscillators corresponding to the normal
modes of the linearized system. \ If $\Lambda_j\ll\delta_i$, for any
$i$ and $j$, then the normal modes are localized, i.e. each normal
mode essentially involves only one of the $\psi_j$; on the contrary,
in the much more interesting case where the $\delta_j$ are of the same
order of magnitude of the $\Lambda_j$, typically the normal modes are
collective motions of the system and, correspondingly, in typical
solutions the term $|\psi_j(t)|^2$ undergoes great changes for each
$j$.

In the opposite limit $\eta\to\infty$ (anticontinuum limit, see \cite{MA}), $\K_0$ becomes a system of decoupled anharmonic oscillators. \ Correspondingly $|\psi_j(t)|^2$ is a constant of
motion. \ One can use KAM or Nekhoroshev theory in order to study the dynamics of $\K$ when $1\ll\eta<\infty$ and to deduce results on the dynamics of the complete NLS equation.

Here we will state a result that can be obtained in this way. \ To
this end, for any $\rho >0$ fixed, we will denote
\bee \Sc^n_\rho:=\left\{(\psi_1,...,\psi_n)\in\C^n\ :\
\sum_{j=1}^{n}|\psi_j|^2=1\ \text{and} \ |\psi_j|>\rho \ ,\ \forall j
\right\} \eee
and we will denote by $\left|\Sc^n_\rho \right|$ its Lebesgue measure.

\begin{theorem}
\label{agg.10}
(KAM theorem $\K$) Consider the Hamiltonian system $\K$, then there
exists a constant $\eta_\star$, such that, for any $|\eta|>\eta_\star$
there exists a set $\Sc_\eta\subset \Sc^n_\rho$ with Lebesgue measure
estimated by
\be
\label{agg.11}
\left|\Sc_\eta\right|\geq \left|\Sc^n_\rho\right|- C\eta^{-1/2}
\ee
with the property that, if the initial datum is in $\Sc_{\eta}$ then
the solution of the Hamiltonian system with Hamiltonian function $\K$
is quasiperiodic and fulfills
\be
\label{agg.12}
\left|\left|\psi_j(t)\right|^2-\left|\psi_j(0)\right|^2\right|<C\eta^{-1/2} 
\ee
for any $t$.
\end{theorem}

To state a corresponding result for the NLS equation, consider the set
$\Le$ of the $\psi\in\X_s$ having $L^2$ norm equal to 1 and fulfilling
\bee
d(\psi,\M)\leq C \exp \left [ -\frac{\mu_*\hbar^{3/2}}{2\mu} \right ] .
\eee
For $\psi\in\Le$, denote $\sum_{j=1}^{n}\tilde \psi_j\hat \varphi_j=\hat\Pi\psi$ and 
\be
\label{agg.14}
\tilde \Sc^n_\rho:=\left\{\psi\in\Le\ :\ |\tilde\psi_j|>\rho \ ,\ \forall j\right\} .
\ee

\begin{corollary}
\label{agg.13}
Consider the NLS equation \eqref{eq1}, under the same assumptions of
theorem \ref{Theorem2} assume also $\eta$ large enough; then there
exists a set $\tilde\Sc_\eta\subset \tilde \Sc^n_\rho$ whose ``measure''
is estimated by
\be
\label{agg.15}
\left|\hat\Pi\tilde\Sc_\eta\right|\geq \left|\hat\Pi\tilde\Sc^n_\rho\right|-  C\eta^{-1/2}
\ee
such that, if $\psi^0\in\tilde\Sc_\eta$ then along the corresponding solution one has
\be
\label{agg.17}
\left|\left|\tilde\psi_j(t)\right|^2-\left|\tilde \psi_j(0)\right|^2\right|<C\eta^{-1/2}
\ee
for all the times $t$ fulfilling \eqref{eq22}.
\end{corollary}

\begin{remark}
\label{agg.18}
All the constants in the above Theorem \ref{agg.10} depend on $n$, in \cite{bam96,BV} one can find some $n$ independent statements. 
\end{remark}

\subsection{The double well potential}\label{s.double} 
In the particular case of a double well potential one can get much more precise results, both for the linear and for the nonlinear system.

By a double well potential we will mean here a potential $V (x)\in C^\infty (\R )$ fulfilling assumptions (i-v) of Section \ref{linear} (with $n=2$) and which moreover is symmetric with respect to spacial reflection $V(-x)=V(x)$. \ It is well known that the {splitting} $\omega$ between the two lowest eigenvalues fulfills the asymptotic estimate
\be
\omega = \frac 12 (\lambda_2-\lambda_1)\le  C e^{-\Gamma/\hbar}, \label {eq722}
\ee
for any $\Gamma < \Gamma_1$ where $\Gamma_1$ is the Agmon distance
between the two wells and $C$ is a positive constant (depending on
$\Gamma$), and moreover the normalized eigenvectors $\varphi_{1,2}$
associated to $\lambda_{1,2}$ can be chosen to be real-valued
functions such that $\varphi_1$ and $\varphi_2$ are respectively of
even and odd-parity, thus, defining the \emph {single well states}
\bee
\varphi_R = \frac {1}{\sqrt 2} \left [ \varphi_1 + \varphi_2 \right ] \ \ \mbox { and } \ \ \varphi_L = \frac {1}{\sqrt 2} \left [ \varphi_1 - \varphi_2 \right ]
\eee
they essentially coincide with the functions $\hat\varphi_j$ used in the multiple well case. \ In particular they fulfill
\be
\| \varphi_{R } \varphi_{L} \|_{L^\infty} \leq Ce^{-\Gamma /\hbar }\ ,\quad \Gamma>0.
\label {eq10}
\ee
Thus, for $\psi\in\Phi_0$, in this section {\it we will write}
\be
\label{agg.20}
\psi=\psi_1\varphi_L+\psi_2\varphi_R
\ee

Here the tunneling gives rise to the so called phenomenon of beating: for almost any initial datum $\psi^0\in\Phi_0$ the expectation value of the position 
\be
\label{eq11}
\langle x \rangle^t = \int_{\R}x\left|\psi^t (x)\right|^2dx
\ee
periodically oscillates between positive and negative values with a period given by $T:=\pi\hbar/\omega$.

In this case the restricted approximate Hamiltonian $\K_0$ takes the form \cite {Sa2}
\be
\label{agg.21}
\K_0:=\bar \psi_2\psi_1+\psi_2\bar\psi_1+\eta\left(|\psi_1|^{2\sigma+2}+ |\psi_2|^{2\sigma+2} \right)
\ee
with the same definition of $\eta$ as in the previous subsections. \
Then Theorems \ref{Theorem2} and \ref{Theorem3} hold together with
their corollaries. \ The main improvement that one can obtain in the
double well case are due to the fact that since the system $\K$ is now
a system with two degrees of freedom with an integral of motion
independent of the Hamiltonian (namely $\Ic$), then it is
integrable. \ This allows us to describe in a very precise way the
trajectories of the system $\K$ which are just the intersection of the
level surfaces of the functions $\K$ and $\Ic$. This is possible since
$\K$ is close to $\K_0$.

{\it To be definite, from now on, we will restrict to the case $\sigma=2$}.

The system $\K_0$, cf. \eqref{agg.21}, has been already studied in \cite {GMS} (see also \cite{RSFS, VA, VA2}) obtaining that, for $|\eta|<2$ almost all solutions perform beating motions, while at $\eta=\pm2$ a bifurcation occurs and new equilibria, localized close to
the minima of the Hamiltonian function, appear. \ As $\eta$ increase the domain of stability of such solutions increase its size, so that, for $\eta$ large enough essentially only localize motions exist. \ Concerning the complete system we can state that if $\eta$ is not at a bifurcation point, then non-homoclinic trajectories associated to the Hamiltonian $\K_0$ approximate the solution $\psi^t$ for times of the order \eqref {eq22}.

\begin{corollary}
\label{Corollary4}
Under the same assumptions as in Theorem \ref{Theorem2}, assume also $\eta\not=\pm 2$;
consider an initial datum such that $\psi_1(0),\psi_2(0)\not=0$, and 
\bee
\K_0(0)\not=\frac{\eta}2-1\ \text{if}\quad \eta>0 \\
\K_0(0)\not=1-\frac{\eta}2\ \text{if}\quad \eta<0
\eee
and also such that $\delta\leq C\mu\hbar^{3/2}$; then, there exists a
positive constant $\mu_\sharp$, depending only on how much the above
quantities differ from the considered values, such that, provided
$\mu<\mu_\sharp\hbar^{3/2}$, there exists a solution of the Hamiltonian system
\eqref{eq17} with trajectory $\gamma$ such that 
\be
\label{eq27}
d(\psi^t,\gamma)\leq C\frac{\mu}{\hbar^{3/2}}\ 
\ee
for the times \eqref{eq26}.
\end{corollary}

\begin{remark}
\label{Remark7}
Homoclinic trajectories are absent when $|\eta |\le 2$; initial conditions such that $\K_0(0) = \frac{\eta}2-1 $, for $\eta >2$, and $\K(0) = 1-\frac{\eta}2$, for $\eta<-2$ corresponds to an homoclinic trajectory. 
\end{remark}

\begin{remark}
The topology of the trajectory $\gamma$ is determined by the condition
\be
\label{eq28}
d(\psi^0,\gamma)\leq C\frac{\mu}{\hbar^{3/2}}\ 
\ee
in the sense that all curves fulfilling this condition have the same topology if the assumptions of the Corollary are fulfilled.
\end{remark}

\begin{remark}
\label{Remark9}
Thus one has that also for the true system beats are present for $|\eta|<2$ while their importance decreases as $|\eta|$ increase above 2. \ In particular for large values of $\eta$ only motions localized close to one well are present \cite{VA2} at least for the time scales
controlled by our theorems.
\end{remark}

\section {Proof of the main results}\label{s.proof}

{\it In order to simplify the notations all the proofs will be carried out in the case of a potential with only 2 wells.}

\subsection{Hamiltonian Formalism}
\label{hamilton}

First consider the real Hilbert space 
\bee
\XR:=D((H_{0,\re})^s)
\eee
where $H_{0,\re}$ is the operator $H_0$ restricted to real valued functions.

We make $\XR\oplus\XR$ a symplectic space by introducing the \emph {semiclassical} symplectic form 
\bee
\alpha\left((p,q);(p',q') \right):=\hbar\int_{\re}\left[p'(x)q(x)-p(x)q'(x) \right] \di x .
\eee
Given a smooth real valued function $\Ha(p,q)$, then we define its Hamiltonian vector field $X_{\Ha}\in \XR\oplus\XR$ by the property
\be
\label{eq48}
\alpha(X_{\Ha},h)= \di\Ha h
\ee
for any $h=(h_p,h_q)\in \XR\oplus\XR$, where $\di\Ha$ denotes the differential of $\Ha$. \ It is well known that $X_{\Ha}$ is in general defined only on a subset of $\XR\oplus\XR$. \ Define also the $L^2$ gradient $\nabla_p\Ha$ of $\Ha$ with respect to $p$ by
\be
\label{eq49}
\left\langle\nabla_p\Ha, h_p\right\rangle_{L^2}:=\int_{\R}\nabla_p\Ha(x) h_p(x)dx  =\di_p\Ha h_p,\quad \forall h_p\in \XR , 
\ee
and similarly we introduce the quantity $\nabla_q\Ha$. \ Then 
\bee
X_{\Ha}=\hbar^{-1}(-\nabla_q\Ha, \nabla_p\Ha),
\eee
and thus the Hamilton equations of $\Ha$ are given by
\be
\label{eq50}
\frac{d}{dt}(p,q)=X_{\Ha}(p,q)\ \iff \ \left ( \dot p=-\frac{1}{\hbar} \nabla_q\Ha\ ,\ \dot q=\frac{1}{\hbar}\nabla_p\Ha \right ) .
\ee
The Poisson brackets between two functions $\Ha$ and $\K$ is defined as
\be
\label{eq51}
\left \{ \Ha,\K \right \} := \alpha (X_\Ha , X_\K ) = - \frac 1\hbar \int_\R \left [ \nabla_p \Ha \nabla_q \K - \nabla_q \Ha \nabla_p \K \right ] \di x 
\ee
which in general is only defined on a subdomain of $\XR \oplus \XR$.

We shall use complex coordinates in $\XR\oplus \XR$ identifying this space with $\X^s$, through
\bee
(p,q)\mapsto \psi=\frac{1}{\sqrt2}(q+\im p) .
\eee 
Therefore, we set 
\be 
\label{eq52} 
\nabla_\psi=\frac{1}{\sqrt2}(\nabla_q-\im\nabla_p) \ \ \mbox { and } \ \ \nabla_{\bar \psi}=\frac{1}{\sqrt2}(\nabla_q+\im\nabla_p) 
\ee
so that, if $\Ha = \Ha (\psi , \bar \psi )$ is a smooth real valued function, we have the identification 
\be 
\label{eq53} 
X_{\Ha}(\psi,\bar \psi)=-\frac{\im}{\hbar} \nabla_{\bar \psi}\Ha(\psi,\bar \psi)\ .
\ee
and in complex coordinates the Poisson brackets are computed by
\be
\label{eq54}
\left\{\Ha,\K\right\}:=\frac{i }{\hbar} \int_{\R} \left [ \nabla_{\psi}\Ha\nabla_{\bar\psi}\K- \nabla_{\bar\psi}\Ha\nabla_{\psi}\K\right ] \di x
\ee

With such a notation then the NLS \eqref{eq1} can be written in the form of a Hamiltonian system, the corresponding Hamiltonian function being the energy eq. \eqref{eq16}.

\begin{remark}
\label{Remark10}
Such a Hamiltonian is invariant under the action of the Gauge group
\be
\label{eq57}
\psi (x) \to \psi (x) e^{-i \beta},
\ee
for any $\beta $ independent of $x$. \ The corresponding conserved quantity is the $L^2$ norm
\be
\label{eq58}
\calN(\psi,\bar \psi)=\int_{\R}|\psi(x)|^2\di x .
\ee
Equivalently one has  
\be
\label{eq59}
\left\{\E,\calN\right\}\equiv0 .
\ee
\end{remark}

Let $\{ \lambda_k \}_{k=1}^\infty$ and $\{ \varphi_k (x) \}_{k=1}^\infty$ be the eigenvalues and the normalized eigenvectors of $H_0$, let 
\be 
\psi (x) = \sum_{k=1}^\infty \zeta_k  \varphi_k (x)\ , 
\label {eq60} 
\ee
and define the Hilbert spaces $\ell^2_s$ of the complex sequences such that
\be
\label{eq61}
\norma{\zeta}_s^2:=\sum_{k\geq1}\lambda_k^s|\zeta_k|^2<\infty
\ee
In such a way we have defined the correspondence
\bee
\psi \in \X^s \leftrightarrow {\mathcal U} (\psi ) :=\zeta \equiv
(\zeta_1,\zeta_2... \zeta_j,...) \in \ell^2_s
\eee
which is a unitary isomorphism. \ In particular, if $\E$ is the Hamiltonian \eqref{eq16}, then $\E\circ\U^{-1}$ (still denoted by $\E$) is the Hamiltonian of the same system written in terms of the new variables $\zeta$. \ In terms of these variables the quadratic part $\E_0$ of the Hamiltonian is given by
\be
\label{eq62}
\E_0=\sum_{j\geq1}\lambda_j\left|\zeta_j\right|^2 
\ee 
and the Poisson brackets can be written as
\bee
\left \{ {\mathcal H} , {\mathcal K} \right \} = \frac i\hbar \sum_{k=1}^{\infty} \left [ \frac {\partial {\mathcal H}}{\partial \zeta_k} \frac {\partial {\mathcal K}}{\partial \bar \zeta_k} - \frac {\partial {\mathcal H}}{\partial \bar \zeta_k}  \frac {\partial {\mathcal K}}{\partial \zeta_k} \right ] 
\eee

\begin {remark} 
From now on we will work in the space $\ell^2_s$ and moreover, in
order to simplify, {\bf we rescale time by the transformation}
$t\mapsto t/\hbar$, thus the Poisson brackets take the form
\be
\left \{ {\mathcal H} , {\mathcal K} \right \} = i \sum_{k=1}^{\infty} \left [ \frac {\partial {\mathcal H}}{\partial \zeta_k} \frac {\partial {\mathcal K}}{\partial \bar \zeta_k} - \frac {\partial {\mathcal H}}{\partial \bar \zeta_k} \frac {\partial {\mathcal K}}{\partial \zeta_k} \right ] \label {eq63}
\ee
\end {remark}

\subsection{Non coupling monomial} 
\label{NonCoupling} 

It is useful to introduce also a different notation for the first two variables (here we recall that we are working in the double well case) and for the remaining ones, thus let us denote
\be
\label{eq64}
u_1:=\zeta_1\ ,\quad u_2:=\zeta_2\ ,\quad z_j:=\zeta_j\ ,\quad j\geq 3 .
\ee

Consider now a monomial of the form 
\be
\label{eq65}
\zeta^K\bar\zeta^L=u^k\bar u^lz^m\bar z^n
\ee
where we used the notations 
\be
\begin {array}{l}
K =(k,m), \ L =(l ,n )  \\ 
k= (k_1,k_2), \ m=(m_3,m_4,m_5,....) ,\ l=(l_1,l_2), \ n=(n_3,n_4,n_5,....) \\
u^k\equiv u_1^{k_1}u_2^{k_2} , \ z^m \equiv z_3^{m_3} z_4^{m_4} \ldots z_q^{m_q} \ldots 
\end {array}
\label{eq66}
\ee

\begin{remark}
\label{Remark12}
A monomial of the form \eqref{eq65} is Gauge invariant, i.e. invariant under the transformation 
\bee \psi\mapsto e^{i\beta}\psi \ \ \mbox { that is } \ \ \zeta_k
\mapsto e^{i \beta} \zeta_k \eee
if and only if $|K|=|L|$, where
\bee
|K|=\sum_j|K_j| .
\eee
In fact \eqref {eq65} is Gauge invariant if, and only if, 
\be
\label{eq67}
0=\left\{\mathcal N(\zeta,\bar\zeta),\zeta^K\bar\zeta^L\right\} = \im \left ( \sum_{j}(K_{j}-L_{j})\right) \zeta^{K}\bar\zeta^{L}
\ee
where $\calN(\zeta,\bar \zeta)=\sum_{j\geq1}|\zeta_j|^2$ is the $L^2$ norm; that is $|K|=|L|$ again.  
\end{remark}

Due to our assumption Hyp.1 on the potential one has 

\begin {lemma}
Let $\sigma (H_0)$ be the spectrum of the linear operator $H_0$ and let $\hbar$ be small enough. \ There exists a sequence of (not necessarily continuous with respect to $\hbar$) functions
$\{E_\gamma(\hbar)\}_{\gamma\in\N}$, $0<E_0 <1$ and there exists a positive constant $C>1$, independent of $\hbar$ and $\gamma$, such that:

\begin{itemize}

\item[i.] $[E_\gamma-C^{-1}\hbar ,E_{\gamma}+C^{-1}\hbar ]\cap \sigma (H_0)=\emptyset$;
 
\item[ii.] $1<E_{\gamma}-E_{\gamma-1}<3$ for all $\hbar$ and all $\gamma\geq1$;

\item [iii.] For any $\hbar$ fixed, we consider the sets of indexes
\bee
J_\gamma (\hbar ) = J_\gamma:=\left\{j\in\N\ :\ E_{\gamma-1}<\lambda_j<E_\gamma  \right\} ,
\eee
then the cardinality of these sets is estimated as 
\bee
\# J_\gamma (\hbar ) \le \frac {C}{\hbar}.
\eee

\end{itemize}

\end {lemma}

\begin {proof} The proof is an immediate consequence of the following result (see Theorem (V-11) in \cite {Rob}, see also Theorem (XIII-81) in \cite {RS4}): 
\bee 
N_{[\alpha ,\beta ]} = \frac {1}{2\pi \hbar} \left [ \left| \left \{ (x,p) \in \R^2 \ : \ \alpha \le p^2 + V(x) \le \beta \right \} \right| + O(\hbar ) \right ] 
\eee
where $N_{[\alpha ,\beta ]} $ is the number of eigenvalues of $H_0$ contained in the interval $[\alpha ,\beta ]$ and where $|.|$ denotes the Lebesgue measure of a set. \ Indeed, let us consider the intervals $[2\gamma , 2 \gamma +1 ]$, $\gamma \in \N$, then the number
of eigenvalues of $H_0$ belonging to these intervals is given by
\bee 
&& N (\gamma) := N_{[2\gamma , 2\gamma +1]} =  \\ 
&& \ = \frac {1}{2\pi \hbar} \left [ \left | \left \{ (x,p) \in \R^2 \ : \ 2\gamma \le
p^2 + V(x) \le 2\gamma +1 \right \} \right | + O(\hbar ) \right ] \le \frac {C}{\hbar} 
\eee
since $V(x) \ge C \langle x \rangle^2$, for some positive constant $C$ independent of $\hbar$ and $\gamma$. \ From this estimate it follows that there exists at least one value $E_\gamma \in (2\gamma , 2 \gamma+1 )$ and $C>1$ satisfying i.. \ Furthermore, conditions ii. and
iii. immediately follow since
\bee
1=(2\gamma )-[2(\gamma -1)+1] \le E_{\gamma}- E_{\gamma -1} \le (2\gamma + 1) - 2(\gamma -1) = 3 
\eee
and
\bee
\# J_\gamma \le N (\gamma ) + N(\gamma -1 ) \le  \frac {2C}{\hbar}. 
\eee
\end {proof}

We fix a sequence with such properties. 

Hereafter, all the perturbative construction will involve only Gauge invariant monomial (hence $|K|=|L|$). \ Keeping this in mind we give the following 

\begin{definition}
\label{Definition2}
A monomial of the form \eqref{eq65} will be called \emph {coupling} if the indexes $(K,L)$ fulfill the following conditions

\begin{itemize}

\item[i.] $|K|=|L|$ 

\item[ii.] $|m|+|n|=1,2$

\item[iii.] if $m_i=1$ and $n_j=1$ then $i\in J_\gamma$ and $j\in J_{\gamma'}$ with $\gamma\not=\gamma'$. 

\end{itemize}

A monomial which is not coupling will be called \emph {non coupling}. \ A polynomial containing {\bf only} coupling (reps. non coupling) monomial will be called coupling (reps. non coupling).
\end{definition}

\begin{remark}
\label{Remark13}
For Gauge invariant monomials the condition {\rm i.} is always
fulfilled. \ Furthermore, in terms of the indexes $k,l,m,n$ condition
{\rm i.} reads
\be
\label{eq68}
|k|-|l|=|n|-|m|
\ee
\end{remark}

\begin{remark}
\label{Remark14}
Any Gauge invariant analytic function can be uniquely decomposed into the sum of a coupling and a noncoupling part. \ We recall that, as in finite dimensional spaces, an analytic function is a function whose Taylor series is convergent (see e.g. \cite{Muj}).
\end{remark}

We also define a new ($\hbar$ dependent) norm in the space $\X^s$ as follows

\begin{definition}
\label{Definition3}
Denote
\be
\label{eq69}
\calN_E(z,\bar z):=\sum_{\gamma}E_\gamma^s\sum_{j\in J_\gamma, j\geq3}|z_j|^2
\ee
then the quantity 
\be
\label{eq70}
\norma{(u,z)}_E^2\equiv \norma{\zeta}_E^2 :=|u_1|^2+|u_2|^2+\calN_E(z,\bar z)
\ee
will be called $E$-norm of $\zeta$. \ Denote $\ell^2_E$ the space of sequences $z=\{ z_j\}_{j \ge 3}$ equipped with the norm $\calN_E(z,\bar z)$. \ By abuse of notation sometime we will also write 
\bee
\norma{z}_E^2:=\calN_E(z,\bar z) .
\eee 
\end{definition}

\begin{remark}
\label{Remark15}
The $E$-norm is equivalent to the standard norm of $\X^s$ with an $\hbar$ independent constant. \ The proof is a trivial computation and is left to the reader. 
\end{remark}

\begin{lemma}
\label{Lemma2}
Let $\zeta^K\bar\zeta^L$ be a non coupling Gauge invariant monomial of degree at most $two $ in $z,\bar z$ (i.e. $|n|+|m| \le 2$), then one has
\be
\label{eq71}
\left \{ \calN_E, \zeta^K \bar\zeta^L \right \}=0 \ \mbox { and } \ \left \{ |u_1|^2+|u_2|^2, \zeta^K\bar\zeta^L \right \}=0
\ee
\end{lemma}

\proof First remark that if $u^k\bar u^lz^n \bar z^m$ is noncoupling of degree at most two in $z,\bar z$ then one has $|k|=|l|$ and thus 
\be
\label{eq72}
\left\{u_1\bar u_1+u_2\bar u_2 , u^k\bar u^lz^n\bar z^m\right\} = i(|k|-|l|) u^k\bar u^lz^n\bar z^m=0 .
\ee
Furthermore, either it is of degree zero in $z,\bar z$ or there exists $\bar \gamma$ and $\bar j,\tilde j\in J_{\bar \gamma}$ such that $n_{\bar j}=1$ and $m_{\tilde j}=1$. \ In the first case one has that 
\bee
\left\{\calN_E, \zeta^K\bar\zeta^L \right\}=\left\{\calN_E (z , \bar z) , u^k\bar u^l \right\} =0
\eee
In the second case, then 
\bee
\left\{\calN_E(z,\bar z) , u^k\bar u^lz^n\bar z^m\right\} &=& u^k\bar u^l \sum_{\gamma} E_\gamma^s \sum_{j\in J_\gamma} \left\{z_j\bar z_j , u^k\bar u^lz^n\bar z^m\right\} \\ 
&=& u^k\bar u^l z^n\bar z^m E_{\bar \gamma}^s \sum_{j\in J_{\bar \gamma}} i(n_j-m_j) =0
\eee
\qed 

Thus the $E$--norm is invariant under the dynamics of a noncoupling Hamiltonian of degree at most $2$ in $z,\bar z$. \ In the more general case one has

\begin{corollary}
\label{Corollary6}
Let $\Ze$ be a non coupling polynomial. Assume that it has a smooth vector field, then there exists $C$ such that
\be
\label{eq73}
\left| \left\{\calN_E,\Ze\right\}(u,\bar u,z,\bar z) \right|\leq C \left[ \calN_E(z,\bar z)\right]^{3/2}
\ee
\end{corollary}

\subsection{Normal form construction} 

Let us rewrite $\E (\psi , \bar \psi )$ as follows
\bee
\E=\Ha_0+\epsilon \Pc_\epsilon
\eee
where
\be
\label{eq74}
\Ha_0(\psi,\bar\psi):=\Omega(|\zeta_1|^2+|\zeta_2|^2)+\sum_{j\geq 3}\lambda_j\left|\zeta_j\right| ^2, \ \ \Omega:=\frac{\lambda_1+\lambda_2}{2} ,
\ee
and
\be
\label{eq75}
\Pc_\epsilon (\psi,\bar\psi):= \frac{\omega}{\epsilon}(|\zeta_2|^2-|\zeta_1|^2)+\Pc_0 (\psi,\bar \psi).
\ee

We are going to prove that there exists a canonical transformation $\Tr_\epsilon$ which gives the Hamiltonian the form
\be
\label{eq76}
\E\circ \Tr_\epsilon=\Ha_0+\epsilon \Ze_\epsilon+\resto
\ee
where $\Ze_\epsilon$ is a non coupling polynomial and $\resto$ has a smooth vector which is exponentially small with $\epsilon^{-1}$.

The construction will be recursive. \ To this end we assume one has been able to construct a canonical transformation $\Tr_r$ putting the Hamiltonian in the form
\be
\label{eq77}
\E\circ \Tr_r\equiv \E^{(r)}=\Ha_0+\epsilon \Ze^{(r)}(\epsilon )+\epsilon^{r+1} \resto^{(r)}(\epsilon) 
\ee
with $\Ze^{(r)}$ being a noncoupling polynomial and where $\resto^{(r)}(\epsilon)$ has a vector field which is bounded, uniformly with respect to $\epsilon$ ($\Ze^{(0)}=0$ and $\resto^{(0)}= \Pc_\epsilon$). \ We look for an auxiliary Hamiltonian ${\G}_{r+1}$ such that considering the corresponding Hamilton equations
\bee
\dot \zeta=X_{{\G}_{r+1}}(\zeta)
\eee
and the corresponding flow $\phi_{r+1}^t$ one has that $\E^{(r)}\circ\phi_{r+1}^{\epsilon^{r+1}} $ is in the form \eqref{eq77} with $r+1$ in place of $r$. 

Explicitly one has
\be
\label{eq78}
\E^{(r)}\circ\phi_{r+1}^{\epsilon^{r+1}}&=&\Ha_0+\epsilon \Ze^{(r)}
\\
\label{eq79}
&+&\epsilon^{r+1}\left(\resto^{(r)}+\left\{\Ha_0,{\G}_{r+1} \right\} \right)
\\
\label{eq80}
&+&\Ha_0\circ \phi_{r+1}^{\epsilon^{r+1}}-\Ha_0-\epsilon^{r+1} \left\{\Ha_0,{\G}_{r+1} \right\}
\\
\label{eq81}
&+&\epsilon\left(\Ze^{(r)}\circ \phi_{r+1}^{\epsilon^{r+1}}- \Ze^{(r)} \right)
\\
\label{eq82}
&+&\epsilon^{r+1}\left(\resto^{(r)}  \circ \phi_{r+1}^{\epsilon^{r+1}}- \resto^{(r)} \right)
\ee
Now, it is quite easy to understand that (\ref{eq80}--\ref{eq82}) are higher order terms (they will be estimated later), while \eqref{eq79} is the term of order $\epsilon^{r+1}$. Thus, if one is able to choose ${\G}_{r+1}$ so that 
\be
\label{eq83}
\resto^{(r)}+\left\{\Ha_0,{\G}_{r+1} \right\}
\ee
is non coupling, then the coupling terms are pushed to order $r+2$, and 
\bee
\Ze^{r+1} =\Ze^{r} + \epsilon^{r} \left [ \resto^{(r)}+\left\{\Ha_0,{\G}_{r+1} \right\} \right ]
\eee
is non coupling.

The main step of the proof is the construction and the estimate of such a function ${\G}$ fulfilling the homological equation associated to \eqref{eq83}.

\subsection{Framework and notations}
Before proceeding to the construction and the estimation of such a ${\G}$ it is useful to extend the setting in which we will work. \ Indeed, since we are working with analytic functions it is useful to work in the \emph {complexification} of our space. \ More precisely, {\it we consider now a phase space in which the variables $v:=\bar u$ and $w:=\bar z$ are independent of $u,z$}. \ Actually this is equivalent to complexify the space in which vary the variables $p,q$
of section \ref{hamilton}. \ With such an extension the Poisson bracket will take the form
\bee
\left \{ \Ha , \K \right \} =i \sum_{k=1}^2 \left [ \frac {\partial \Ha }{\partial u_k} \frac {\partial \K}{\partial v_k} - \frac {\partial \K}{\partial u_k} \frac {\partial \Ha}{\partial v_k} \right ] + i \sum_{j=3}^\infty \left [ \frac {\partial \Ha}{\partial z_j} \frac {\partial \K}{\partial w_j}- \frac {\partial \K}{\partial z_j} \frac {\partial \Ha}{\partial w_j} \right ]
\eee
Such a phase space will be denoted by $\X^s_\C$. In $\X^s_\C$ we will use the norm
\be
\label{eq84}
\lceil (u,v,z,w) \rceil :=\max\left\{\norma{(u,v)}_{\C^2},\norma{(z,w)}_E \right\}  
\ee
with $\norma{(z,w)}_E$ defined by
\be
\label{eq85}
\norma{(z,w)}_E^2:=\norma{z}_E^2+\norma{w}_E^2, \ \mbox { where }\ \| \cdot \|_E^2 =\calN_E (\cdot , \bar {\cdot }) .
\ee
{\it Sometimes it is also useful to use more compact notations for the phase space variables, thus we will also write $\zeta:=(u,z)$, $\eta:=(v,w)$, $\xi:=(\zeta,\eta)$.}

Moreover, given $R>0$ we will denote by $B_R$ the ball of radius $R$ and center 0 in the phase space:
\bee
B_R = \left \{ \xi \in \X^s_\C \ : \ \lceil \xi \rceil \le R \right \}.
\eee
Given an analytic function $\Ha$ with Hamiltonian vector $X_{\Ha}$ analytic as a map from $B_R$ to $\X^s_\C$ we will denote
\be
\label{eq86}
|X_{\Ha}|_R:=\sup_{\xi \in B_R}\lceil X_{\Ha}(\xi)\rceil \ \ \mbox { and } \ \ |\Ha|_R:=\sup_{\xi \in B_R}|\Ha(\xi)|\ ,
\ee
with norm defined by \eqref{eq84}.

\begin{remark}
\label{Remark16}
By Corollary \ref {Corollary1}, our initial perturbation $\Pc_\epsilon$, defined by \eqref {eq75}, is such that
\be
\label{eq87}
\normag{\Pc_\epsilon}R\leq \frac{\omega}{\epsilon}R+\frac{C}{\hbar^\sigma}R^{2\sigma+1} 
\ee
\end{remark}

\subsection{Solution of the Homological equation}
\label{homological}

In this section we will construct and estimate the solution of the Homological equation associated to \eqref{eq83}; that is we look for $\G_{r+1}$ such that \eqref {eq83} is non coupling. 

Precisely, consider the Gauge invariant Hamiltonian function $\Ha$ and decompose it (Remark \eqref {Remark14}) as $\Ha=\F+\Ze$ with $\Ze$ non coupling and $\F$ coupling, and consider the equation
\be
\label{eq88}
\left\{\Ha_0 ,\G\right\}=\F , \ \ \F = \mbox { coupling part of } \Ha
\ee
then we are going to prove the following.

\begin{theorem}
\label{Theorem5}
If $\Ha$ has a Hamiltonian vector $X_{\Ha}$ analytic as a map from
$B_R$ to $\X^s_\C$, then there exists a solution $\G$ of \eqref{eq88}
which is coupling and has a vector field $X_\G$ with the same analyticity
properties; moreover there exists a positive $\alpha=\alpha(\hbar)$
such that the following inequality holds
\be
\label{eq89}
\normag{\G}R\leq\frac{1}{\alpha}\normag{\Ha}R
\ee
moreover $\alpha(\hbar)\geq C^{-1}\hbar^{3/2}$
\end{theorem}

Before starting the proof we need some preparation. Since $\F$ is
coupling then it is of degree at most $2$ with respect to $z$ and
$w$. \ Therefore, we can decompose $\F$ as follows
\be
\label{eq90}
\F:=\F^{10}+\F^{01}+\F^{20}+\F^{11}+\F^{02}
\ee
with 
\be
\label{eq91}
\F^{10}=\left\langle F^{10}(u,v);z \right\rangle,\quad 
\F^{01}=\left\langle F^{01}(u,v);w \right\rangle 
\\
\label{eq92}
\F^{20}=\left\langle F^{20}(u,v)z;z \right\rangle ,\quad 
\F^{02}=\left\langle F^{02}(u,v)w;w \right\rangle
\\
\label{eq93}
\F^{11}=\left\langle F^{11}(u,v)z;w \right\rangle
\ee
and $\left\langle.;.\right\rangle$ be the scalar product of the real Hilbert space $\ell^2$ defined by
\be
\label{eq94}
\left\langle z^{(1)};z^{(2)}\right\rangle:=\sum_{j\geq 3}z^{(1)}_j z^{(2)}_j 
\ee
and $F^{01},F^{10}$ which are $\X^s_{\C}$ valued functions of $(u,v)$ and $F^{20},F^{02},F^{11}$ functions taking values in suitable spaces of linear operators. \ We will denote by $F^{01}_j(u,v)$ the components of $F^{01}$; we will denote by $F^{02}_{ij}(u,v)$ the
matrix elements of $F^{02}(u,v)$, and similarly for the other quantities.

\begin{lemma}
\label{Lemma3}
The Homological equation \eqref{eq88} has a formal solution $\G$ with the same structure as $\F$ (cf. \eqref{eq90}-\eqref{eq93}), but with the quantities $G$'s defined by
\be
\label{eq95}
G^{10}_j(u,v):=\frac{F^{10}_j(u,v)}{i(\lambda_j-\Omega)},\quad
G^{01}_j(u,v):=\frac{F^{01}_j(u,v)}{-i(\lambda_j-\Omega)}
\\
\label{eq96} 
G^{20}_{jl}(u,v):=\frac{F^{20}_{jl}(u,v)}{i(\lambda_j+\lambda_l-2\Omega)}
,\quad
G^{02}_{jl}(u,v):=\frac{F^{02}_{jl}(u,v)}{-i(\lambda_j+\lambda_l-2\Omega)}
\\
\label{eq97}
G^{11}_{jl}(u,v):=\frac{F^{11}_{jl}(u,v)}{i(\lambda_j-\lambda_l)}
\ee
\end{lemma}

\begin{remark}
\label{Remark17}
In \eqref{eq97} one always has $j\not=l$ since terms of the form $F^{11}_{jj}(u,v)$ are always noncoupling; furthermore, one has that $j$ and $l$ cannot belong to the same set $J_\gamma$. \ The solution $\G$ is \emph {coupling}.
\end{remark}

\proof We consider explicitly only the term $\G^{01}$, since all the other terms can be studied exactly in the same way. \ First define 
\bee
\Ha^{(1)}(u,v):=\Omega(u_1v_1+u_2v_2)\ ,\quad \Ha^{(\infty)}(z,w):=\sum_{j\geq3} \lambda_jz_jw_j , 
\eee
so that $\Ha_0=\Ha^{(1)}+\Ha^{(\infty)}$. \ Now one has
\be
\label{eq98}
\left\{\Ha_0,\G^{01}\right\} &=&
\left\{\Ha^{(1)},\G^{01}\right\}+\left\{\Ha^{(\infty)},\G^{01}\right\}
\\ &=& \sum_{j}\left[\left\{\Ha^{(1)},G_j^{01} \right\}-i\lambda_j
G^{01}_j \right]w_j \nonumber \ee
where we simply computed explicitly $\left\{\Ha^{(\infty)},\G^{01}\right\}$. \ Thus to ask
\be
\label{eq99}
\left\{\Ha_0,\G^{01}\right\}=\F^{01} 
\ee
is equivalent to ask
\be
\label{eq100}
\left\{\Ha^{(1)},G^{01}_j \right\}-i\lambda_j G^{01}_j=F^{01}_j
\ee
We compute now the above Poisson Bracket. \ To this end decompose $G^{01}_j$ in Taylor series, one has 
\be
\label{eq101}
G^{01}_j(u,v)=\sum_{kl}G^{01}_{j,kl}u^kv^l\ ,
\ee
from which 
\be
\label{eq102}
\left\{\Ha^{(1)},G^{01}_j \right\}=\sum_{kl}G^{01}_{j,kl} i\Omega (|l|-|k|)u^kv^l 
\ee
but, assuming that $\G^{01}$ is coupling (which will be verified in a while), due to the limitation \eqref{eq68}, one has $|k|-|l|=|m|-|n|=-1$ (due to the fact that $\G^{01}$ is linear in $w$), and therefore
\be
\label{eq103}
\left\{\Ha^{(1)},G_j^{01} \right\}=i\Omega G_j^{01}\ .
\ee
Inserting this expression in \eqref{eq100} one gets
\bee
i\Omega G_j^{01}-i\lambda_j G^{01}_j=F^{01}_j
\eee
which shows that the function $\G^{01}$ of \eqref{eq95} actually fulfills the equation \eqref{eq99} and is coupling. \ All the other terms can be studied in the same way; a detailed proof is omitted.
\qed

\noindent
{\it End of the Proof of Theorem \ref{Theorem5}}. \ We will explicitly
prove only the estimate of the norms of the vector fields of
$X_{\G^{01}}$ and $X_{\G^{11}}$, the other being similar and
simpler. \ We start with $X_{\G^{01}}$. \ Consider first the vector
field of $\F^{01}$, whose components have the form
\be
\label{eq104}
X^{u_l}_{\F^{01}}=i\left\langle\frac{\partial F^{01}}{\partial v_l}; w \right\rangle 
\\
\label{eq105}
X_{\F^{01}}^z =iF^{01}
\ee
Since $\F$ is the coupling part of $\Ha$ then
\be
\label{eq106}
X^{u}_{\F^{01}}=d_wX_{\Ha}^{u}(u,v,0,0)w
\ee
and therefore, adding also the $v$ components, Cauchy inequality implies that
\be
\label{eq107}
\left \|
\frac{\partial F^{01}}{\partial (u,v)}(u,v)\right \|_{{\mathcal L}
(\X^s_{\C},\C^4)}\leq \frac{1}{R} \sup_{B_R} \lceil {X_{\Ha}^{(u,v)} (u,v,z,w)} \rceil = \frac 1R \left | X_{\Ha}^{(u,v)} \right |_R
\ee
where the norm at l.h.s. is the norm as a linear operator from $\X^s_{\C}$ to $\C^4$. \ Using this inequality and making use of \eqref {eq95} it is very easy to estimate $X^{(u,v)}_{\G^{01}}$: to this end define $\tilde w_j:=w_j/i(\lambda_j-\Omega)$ and let $C>1$
be such that
\be
\label{eq108}
C^{-1} \hbar:=\inf_{j\geq 3}|\lambda_j-\Omega|.
\ee
Then one has 
\bee
\norma{\tilde w}_E\leq C\norma  w_E / \hbar
\eee
and, using the definition of $\G^{01}$, cf. \eqref{eq95}, one has 
\be
\label{eq109}
X^{u}_{\G^{01}} =i\left\langle\frac{\partial F^{01}}{\partial v}; \tilde w \right\rangle 
\ee
and similarly for the $v$ components, and thus
\be
\label{eq110}
\left | {X_{\G^{01}}^{(u,v)}} \right |_R \leq \frac{C}{\hbar} \left | {X_{\Ha}^{(u,v)}} \right |_R
\ee
for some $C>1$. \ The estimate of the $z$ components of the vector field is simpler (due to the simpler form of the $z$ component) and collecting the two one gets
\be
\label{eq111}
\left | {X_{\G^{01}}} \right |_R \leq \frac{C}{\hbar} \left | {X_{\Ha}} \right |_R 
\ee

We come now to the estimate of the vector field of $\G^{11}$. \ Preliminary to this estimate we remark that the components of the vector field of $\F^{11}$ are given by
\be
\label{eq112}
X_{\F^{11}}^{(u,v)}=d^2_{zw}X_{\Ha}^{(u,v)}(u,v,0,0)[z,w]
\\
\label{eq113}
X_{\F^{11}}^{w}=d_{z}X_{\Ha}^{w}(u,v,0,0)z=F^{11}z
\ee
thus in particular, by Cauchy inequality,
\be
\label{eq114}
\sup_{B_R}\norma{F^{11}(u,v)}_{ {\mathcal L} (\ell^2_E,\ell^2_E)} \le \frac{1}{R} \left | {X_{\Ha}^{w}} \right |_R
\ee
where the norm at l.h.s. is the norm as a linear operator from $\ell^2_E$ to $\ell^2_E$.

In order to estimate $X^z_{\G^{11}}$ we have just to estimate the norm of the operator (from $\ell^2_E$ to itself) whose matrix is defined in \eqref{eq97}. \ To this end remark that the boundedness of $G^{11}$ as an operator from $\ell_E^2$ to itself (or to its dual) is
equivalent to the boundedness of the operator with matrix elements $t_l G^{11}_{jl}s_j$ as an operator from $\ell^2$ to itself, where $t_l,s_j$ are suitable positive numbers (in fact, $\ell^2$ and $\ell^2_E$ are spaces with not equivalent norms). \ Thus Lemma \ref{Lemma7} of the
appendix ensures that
\be
\label{eq115}
\sup_{B_R}\norma{G^{11}(u,v)}_{{\mathcal L} (\ell^2_E,\ell^2_E)} \le \frac{1}{\alpha } \left | {X_{\Ha}^{w}} \right |_R , \ \ \alpha \ge C^{-1}\hbar^{3/2},
\ee
for some $C>1$, since $|\lambda_j - \lambda_l |\ge C^{-1}\hbar$ if $j$ and $l$ belong to different sets $J_\gamma$. \ Working in a similar way for the other components and the other parts of the vector field of the function $\G$ one gets the result. \qed

From the proof (especially \eqref {eq106}, \eqref {eq112} and \eqref {eq113}) also the following useful Lemma follows

\begin{lemma}
\label{Lemma4}
The following estimates hold
\be
\label{eq116}
\normag{\F}R\leq 5\normag{\Ha}R\ ,\quad \normag{\Ze}R\leq 6\normag{\Ha}R \ ,\quad \left|\Ze\right|_R\leq 6\left|\Ha \right|_R
\ee
\end{lemma}

\subsection{Quantitative estimates}
\label{quantitative}

First we fix a positive $R$ in such a way that $X_{\Pc_\epsilon}$ is analytic on $B_R$, and choose constants $P$, $P^*$ (that depend on $\hbar$ and on all the other parameters) such that
\be
\label{eq117}
\normag{\Pc_\epsilon}R\leq P\ ,\quad \left|\Pc_\epsilon\right|_R\leq P^* .  
\ee
All along this section we fix a small value of $\hbar$, and will make explicit estimates so that at the end it will be possible to insert the dependence on $\hbar$ of the final estimate.

\begin{lemma}
\label{Lemma5}
(Iterative Lemma) Consider a Gauge invariant Hamiltonian of the form 
\be
\label{eq118}
\E^{(r)}=\Ha_0+\epsilon \Ze^{(r)}+\epsilon^{r+1}\resto^{(r)} ,
\ee
with $\Ze^{(r)}$ non coupling and where $\resto^{(0)}= \Pc_\epsilon $ and $\Ze^{(0)}=0$. \ Fix $\delta<R/(r+1)$, assume that the Hamiltonian vector fields of $\Ze^{(r)}$ and of $\resto^{(r)}$ are analytic on $B_{R-r\delta}$, and that 
\be
\label{eq119}
&&\normag{\Ze^{(r)}}{R-r\delta}\leq\left\{ 
\begin{matrix}
0 &{\rm if} & r=0
\\
6P\sum_{i=0}^{r-1} \displaystyle{\mus} ^i & {\rm if} &r\geq1
\end{matrix}
\right. 
\ ,
\\
&& \left|{\Ze^{(r)}}\right|_{R-r\delta}\leq\left\{ 
\begin{matrix}
0 &{\rm if} & r=0
\\
6P^*\sum_{i=0}^{r-1} \displaystyle{\mus} ^i & {\rm if} &r\geq1
\end{matrix}
\right. 
\\
&&\label{eq121}
\normag{\resto^{(r)}}{R-r\delta}\leq \frac P{\epsilon_0^r}\ ,\quad
\left|{\resto^{(r)}}\right|_{R-r\delta}\leq \frac{P^*}{\epsilon_0^r}\ , 
\ee
with
\be
\label{eq122}
\epsilon_0:=\frac{\alpha\delta}{75P} ,\ \ \mbox { where } \ \ \alpha = \alpha (\hbar ) \ge C^{-1}\hbar^{3/2},
\ee
then, if $\displaystyle{\mus} <1/2$ there exists a Hamiltonian function $\G_{r+1}$ analytic on $B_{R-\delta r}$ generating the canonical transformation $\phi_{r+1}^{\epsilon^{r+1}}$ such that $\E^{(r)}\circ\phi^{\epsilon^{r+1}}_{r+1}$ has the form \eqref{eq118} and satisfies the estimates \eqref{eq119}-\eqref{eq121} with $r+1$ in place of $r$, moreover the new Hamiltonian is Gauge invariant and the canonical transformation fulfills
\be
\label{eq123}
\sup_{\lceil \xi \rceil \le R-(r+1)\delta} \left \lceil {\xi-\phi^{\epsilon^{r+1}}(\xi)} \right \rceil \le \epsilon \mus^r \frac{P}{\alpha}
\ee
\end{lemma}
\proof Decompose $\resto^{(r)}$ into its coupling part $\F_r$ and its noncoupling part $\Ze_r$. \ Define $\Ze^{(r+1)}:=\Ze^{(r)}+\epsilon^r\Ze_r$ and use Lemma \ref{Lemma4} to estimate $\normag{\Ze^{(r+1)}}{R-\delta(r+1)}$. \ Use Theorem \ref{Theorem2} to construct $\G_{r+1}$ as the solution of
\bee
\left\{\Ha_0;\G_{r+1}\right\}=-\F_r
\eee
then 
\be
\label{eq124}
\normag{\G_{r+1}}{R-r\delta}\le \frac 1\alpha \left | X_{\resto^{(r)}} \right |_{R-r\delta} \le \frac P{\alpha\epsilon_0^r}
\ee
The Hamiltonian $\E^{(r)}\circ \phi^{\epsilon^{r+1}}$ was computed in subsection \ref{NonCoupling} and is given by equations \eqref{eq78}--\eqref{eq82}, which has the form \eqref{eq118} provided one defines
\bee
\resto^{(r+1)}=\resto^{(r+1)}_a+\resto^{(r+1)}_b+\resto^{(r+1)}_c
\eee
where 
\bee
\resto^{(r+1)}_a = \epsilon^{-(r+2)}\eqref{eq80}, \ \resto^{(r+1)}_b = \epsilon^{-(r+2)}\eqref{eq81} \ \mbox { and } \ \resto^{(r+1)}_c = \epsilon^{-(r+2)}\eqref{eq82}.
\eee
Then, from Lemma \ref{Lemma10}, with $\mu = \epsilon^{(r+1)}$, it follows that
\bee
\left | X_{\resto^{(r+1)}_b} \right |_{(R-r\delta ) -\delta } & \le & \frac {5}{\delta \epsilon^{r+1}} \mu \left | X_{\Ze^{(r)}} \right |_{R-r\delta }\left | X_{\G_{(r+1)}} \right |_{R-r\delta} \\ 
&\le & \frac {5}{\delta} \left [ 6 P \sum_{i=0}^{r-1} \left ( \frac {\epsilon}{\epsilon_0} \right )^i \right ] \left ( \frac {P}{\alpha \epsilon_0^r} \right )\\
& \le & \frac {60 P^2}{\alpha \delta \epsilon_0^r}
\eee
since ${\epsilon} \le \frac 12 \epsilon_0$. \ Similarly
\bee
\left | X_{\resto^{(r+1)}_c} \right |_{(R-r\delta ) -\delta } & \le & \frac {5}{\delta \epsilon} \mu \left | X_{\resto^{(r)}} \right |_{R-r\delta }\left | X_{\G_{(r+1)}} \right |_{R-r\delta} \\ 
&\le & \frac {5\epsilon^r}{\delta} \left ( \frac {P}{\epsilon_0^r} \right ) \left ( \frac {P}{\alpha \epsilon_0^r} \right ) \le \frac {5 P^2}{\alpha \delta \epsilon_0^r}\left ( \frac {\epsilon}{\epsilon_0} \right )^r \\ 
&\le & \frac {5 P^2}{2^r \alpha \delta \epsilon_0^r}
\eee
Furthermore, from Lemma \ref{Lemma11}, with $\Ha = \resto^{(r)}$ and $\mu = \epsilon^{(r+1)}$, it follows that 
\bee
\left | X_{\resto^{(r+1)}_a} \right |_{(R-r\delta ) -\delta } & \le & \frac {25}{\delta \epsilon^{r+2}} \mu^2 \left | X_{\resto^{(r)}} \right |_{R-r\delta }\left | X_{\G_{(r+1)}} \right |_{R-r\delta} \\ 
&\le & \frac {25\epsilon^r}{\delta} \left ( \frac {P}{\epsilon_0^r} \right ) \left ( \frac {P}{\alpha \epsilon_0^r} \right ) \le \frac {25 P^2}{\alpha \delta \epsilon_0^r}\left ( \frac {\epsilon}{\epsilon_0} \right )^r \\ 
&\le & \frac {25 P^2}{2^r \alpha \delta \epsilon_0^r}
\eee
Hence
\be
\left | X_{\resto^{(r+1)}} \right |_{(R-r\delta ) -\delta } &\le & \frac {P}{\alpha \delta} \left [ \frac {25 P}{2^r \epsilon_0^r} + \frac {60P}{\epsilon_0^r} + \frac {5 P}{2^r \epsilon_0^r} \right ] \nonumber \\ 
&\le & \frac {P}{\alpha \delta} \left [ \frac { P}{\epsilon_0^r} \left ( 60 + \frac {30}{2^r} \right ) \right ] \label{eq125}
\ee
If $r=0$ then the second term is not present and the square bracket in \eqref{eq125} is equal to $30 P$; if $r\geq 1$, one has that the square bracket in \eqref{eq125} is less than $75P/\epsilon_0^r$ and thus the thesis on the vector fields follows. \ The estimates of the moduli are obtained in a similar way from Lemma \ref {Lemma4}. \qed

It is clear that the Hamiltonian \eqref{eq16} of the NLS fulfills the assumptions of the Lemma with $r=0$ and thus the Lemma allows us to put our Hamiltonian in normal form up to any order $r$. \ To obtain the exponentially small estimate of the remainder take $\delta=R/2r$ in order to fix the domain of definition of the final Hamiltonian, and then choose an optimal value of $r$, this can be done by minimizing the estimate of the remainder or simply choosing
\bee
r_*:=\left[ \frac{R \alpha}{150 e P \epsilon} \right]
\eee
(with $[.]$ denoting the integer part). \ One thus obtains the following Theorem which is the main technical result of the paper, where $\Ze = \epsilon \Ze^{(r_\star)}$ and $\resto = \epsilon^{r_\star +1} \resto^{(r_\star )}$. 

\begin{theorem}
\label{Theorem6}
Consider the Hamiltonian $\E$ (cfr. \eqref{eq16}), define 
\be
\label{eq126}
\epsilon_*:=\frac{\alpha R}{150 eP} ,
\ee
assume $|\epsilon|<\epsilon_*/2$ then there exists an analytic canonical transformation $\Tr$ such that 
\be
\label{eq127}
\E\circ\Tr =\Ha_0+ \Ze+\resto ,
\ee
with $\Ze$ non coupling; both $\Ze$ and $\resto$ have a vector field which is analytic in $B_{R/2}$ and fulfill the estimate 
\be
\label{eq128}
\normag{\Ze}{R/2}\leq 12 \epsilon P\ ,\quad\normag{\resto}{R/2}\leq \epsilon P \exp \left [ -\left(\frac{\epsilon_*}{\epsilon} \right) \right ] \\
\label{eq129}
\left|\Ze\right|_{R/2}\leq 12 \epsilon P^* ,\quad\left|{\resto}\right|_{R/2}\leq \epsilon P^* \exp \left [ -\left(\frac{\epsilon_*}{\epsilon} \right) \right ]
\ee
Moreover the transformed Hamiltonian is Gauge invariant and the canonical transformation fulfills
\be
\label{eq130}
\sup_{\lceil  \xi \rceil \leq R/2} \lceil {\xi-\Tr(\xi)}\rceil \leq
2\epsilon \frac{P}{\alpha}
\ee
\end{theorem}

It is also important to reformulate the Theorem computing $P$ in terms
of the original quantities $\omega,\epsilon, \hbar$ i.e. using
\eqref{eq87}, thus one easily gets the following

\begin{corollary}
\label{Corollary7}
Fix a positive $R$, consider the Hamiltonian $\E$  (cfr. \eqref{eq16}), define the small parameter
\be
\label{eq131}
\mu:=\omega+\frac{|\epsilon|}{\hbar^\sigma} ,
\ee
then there exists $\mu_*>0$ independent of $\hbar,\omega,\epsilon$, such that, if $\mu<\mu_*\hbar^{3/2}/2$ then there exists an analytic canonical transformation $\Tr$ which transform $\E$ into \eqref {eq127}, where  
\be
\label{eq132}
\normag{\Ze}{R/2}\leq C\mu \ ,\quad\normag{\resto}{R/2}\leq C\mu
\exp \left [ -\frac{\mu_*\hbar^{3/2}}{\mu} \right ]
\\
\label{eq133}
\left|\Ze\right|_{R/2}\leq C\mu ,\quad\left|{\resto}\right|_{R/2}\leq C\mu
\exp \left [ - \frac{\mu_*\hbar^{3/2}}{\mu} \right ] 
\ee
Moreover the transformed Hamiltonian is Gauge invariant and the canonical transformation fulfills
\be
\label{eq134}
\sup_{\lceil  \xi \rceil \leq R/2} \lceil {\xi-\Tr(\xi)}\rceil \leq C\frac{\mu}{\hbar^{3/2}} 
\ee
\end{corollary}
Theorem \ref {Theorem3} is a direct corollary of Corollary \ref{Corollary7}.

The manifold $z=0$ is approximately invariant for the dynamics of the
system \eqref{eq127} and on such a manifold the dynamics is that of a
Hamiltonian system with a Hamiltonian function which is an exponentially
small perturbation of
\be
\label{eq135}
\K(u,\bar u):=\Ze(u,\bar u,0,0) .
\ee
Such a system has the additional integral of motion 
\be
\label{eq136}
\Ic(u,\bar u)=\sum_{j}|u_j|^2 \ .
\ee

In the true nonlinear system one has

\begin{theorem}
\label{Theorem7} Assume $\mu<\mu_* \hbar^{3/2}$ and consider the
Cauchy problem for the system \eqref{eq127} which is equivalent to
NLS. \ Define
\bee
\delta^2=\max\left\{\exp \left [ - \frac{\mu_*\hbar^{3/2 }}{\mu }\right ], \norma{z_0}_s ^2\right\}
\eee
($z_0$ being the initial datum for $z$) then, one has
\be
\label{eq137}
\norma{z(t)}_s\leq C\delta 
\ee
and
\be
\label{eq138}
\left|\Ic(t)-\Ic(0)\right|\leq C\delta^2\ ,\quad \left|\K(t)-\K(0)\right|\leq C\mu\delta^2
\ee
for all times $t$ fulfilling
\be
\label{eq139}
|t|\leq \frac{1}{C\mu\delta}.
\ee
\end{theorem}

\proof First remark that due to the equivalence of the $E$ norm and the $s$ norm one has 
\be
\sqrt{\calN_E (\zeta_0)}=\norma{z_0}_E\leq C\delta. 
\ee 
Thus, by \eqref {eq73} and \eqref {eq133} one has
\be
\label{eq141}
\frac {d \calN_E}{dt }\leq C\mu(\calN_E)^{3/2}+C\mu \exp \left [- \frac{\mu_*\hbar^{3/2}}{\mu }\right ] \sqrt{\calN_E}\leq C\mu(\calN_E)^{3/2} 
\ee
which can be solved giving the estimate 
\be
\label{eq142}
\calN_E(t)\leq 2\delta^2 
\ee
for the times \eqref{eq139}. \ The estimate of $\norma{z(t)}_s$ follows from equivalence of the norms. To obtain the estimate of the diffusion of $\Ic$ simply remark that $\calN=\Ic+\calN_E$ is an exact integral of motion and use the estimate \eqref{eq142}. \ Finally, to estimate the diffusion of $\K$ remark that the total Hamiltonian is an integral of motion, but one has
\be
\label{eq143}
\left|\E\circ\Tr-\Ha_0+\K\right|\leq C\mu\left(\norma{z(t)}_s^2+\exp \left [ -\frac{\mu_*\hbar^{3/2}}{\mu }\right ] \right)
\ee
which using the previous estimates implies the thesis.\qed

\noindent
{\it Proof of Theorem \ref{Theorem2}. } Define first $\M:=\Tr(\Phi_0)$. \ In order to distinguish between the original coordinates of the NLS equation and the new coordinates introduced by the transformation $\Tr$ we will denote the new variables by adding a prime, i.e. we will write $\zeta=\Tr(\zeta')$. \ Remark that since $\Tr$ is Lipschitz together with its inverse, then for any couple of points $\zeta,\tilde \zeta$ one has  
\begin{equation}
\label{lip.1}
C^{-1}\norma{\zeta-\tilde \zeta}_s\leq \norma{\zeta'-\tilde \zeta'}_s
\leq C \norma{\zeta-\tilde \zeta}_s
\end{equation}
and therefore 
\begin{equation}
\label{lip.2}
d(\psi^0,\M)\leq \delta\Longrightarrow d(\zeta^0\null',\Phi_0)\leq
C\delta \iff \norma{z_0'}_s\leq C\delta
\end{equation}
where by a slight abuse of notation we wrote $\psi^0=\Tr(\zeta'
)$. \ Thus eq. \eqref{eq137} implies \eqref{eq23} for the considered
times. \ To get \eqref{eq24}, just remark that (again with a slight
abuse of notation)
\bee
\Pi_c\psi=z=z'+\asy\left(\frac{\mu}{\hbar^{3/2}}\right) 
\eee
and thus \eqref{eq137} implies \eqref{eq24}.\qed

The proof of Corollary \ref{Corollary3} is also a direct consequence
of eqs. \eqref{eq138} when one considers the deformation induced by
the change of variables. A detailed proof is omitted. 

\appendix

\section{Proof of Theorem \ref{Theorem1}}
\label{appendiceA}

For possible future reference, in this section we will work in $\R^d$.

We start by recalling some notations and definitions from Robert \cite{Rob} that will be used in the following

\begin{definition}
\label{Definition4}
A function $m:\re^{2d}\to[0,+\infty)$ is called a \emph {tempered weight} if there exist $C_0,N_0>0$ such that
\be
\label{eq144}
m(X)\leq C_0m(X_1)(1+|X-X_1|)^{N_0}\ ,\quad \forall X,X_1\in\re^{2d}
\ee
\end{definition}

\begin{definition}
\label{Definition5}
A function $a\in C^\infty(\re^{2d})$ is called a semiclassical symbol with weight $(m,\rho)$, with $\rho\geq 0$, if for any multi index $\alpha$ there exists a constant $C_\alpha$ such that 
\be
\label{eq145}
\left|\partial^{\alpha}a(X)\right|\leq C_\alpha \frac{m(X)}{(1+|X|)^{\rho|\alpha|}} \ ,\quad \forall X\in\re^{2d} 
\ee
In this case we write $a\in\Sigma^{m}_\rho$.
\end{definition}

In the following we will also need an extension to $\hbar$ dependent families of symbols.

\begin{definition}
\label{Definition6}
A smooth map $\hbar\mapsto a(\hbar)\in \Sigma^m_\rho$ is called an $\hbar$-admissible symbol of class $(m,\rho)$, if for any integer $N$ large enough one has
\be
\label{eq146}
a(\hbar)=a_0+\hbar a_1+\hbar^2a_2+...+\hbar^Na_N+r_{N+1}
\ee
with $a_j\in\Sigma^{m,-2j}_\rho$ and $r_{N+1}(\hbar)$ bounded in $\Sigma_\rho^{m,-2(N+1)}$. Here $\Sigma^{m,K}_\rho$ is the class of symbols with weight
\bee
\left(m\, (1+|X|)^{K\rho},\rho\right).
\eee 
\end{definition}

Given an $\hbar$-admissible symbol $a$ of class $(m,\rho)$, one defines the corresponding Weyl operator acting on $L^2$ by
\be
\label{eq147}
Op^w(a)\psi(x):=\frac{1}{(2\pi\hbar)^d}\int_{\re^{2d}}\e^{\im \frac{(x-y,\xi)}{2}}a\left(\frac{x+y}{2},\xi,\hbar \right)\psi(y)\mbox {\rm d}^dy \mbox {\rm d}^d\xi
\ee
By the theory of \cite{Rob} one has that, for any $\hbar>0$, such an operator is well defined on the Schwartz space. \ Under suitable conditions it extends to a selfadjoint operator on $L^2$.

\begin{definition}
\label{Definition7}
A strongly admissible operator of weight $(m,\rho)$ is a $C^{\infty}$ application
\bee
A:(0,\hbar_*)\to \Le(S(\re^d),L^2(\re^d))
\eee
such that there exists an admissible symbol $a\in\Sigma_\rho^m$ such that $A(\hbar)= Op^w(a(\hbar))$.
\end{definition}

One of the most important properties of strongly admissible operators is the given by the following Theorem. 

\begin{theorem}
\label{Theorem8}
(Theorem II-32 of \cite{Rob}) Let $A=Op^w(a)$ and $B=Op^w(b)$ be two
strongly admissible operators of weights $(m,\rho)$ and $(n,\rho)$
respectively; then $AB$ is a strongly admissible operator with symbol
$c(\hbar)$ and weight $(mn,\rho)$. \ Moreover, if
$a\sim\sum_{j\geq0} \hbar^ja_j$ and $b\sim\sum_{j\geq0} \hbar^jb_j$
then $c\sim\sum_{j\geq0} \hbar^jc_j$ with
\be
\label{eq148}
c_j=\sum_{|\alpha|+|\beta|+k+l=j}\frac{1}{\alpha!\beta!} \left(\frac{1}{2}\right)^{|\alpha|} \left(\frac{1}{2}\right)^{|\beta|} (\partial_\xi^{\alpha}D_x^{\beta}a_k)  (\partial_\xi^{\beta}D_x^{\alpha}b_l) 
\ee
where $D_x=-\im \partial_x$. \ Furthermore, for any $N\geq 0$, one has 
\be
\label{eq149}
c=\sum_{j=0}^{N}\hbar^jc_j+\hbar^{N+1}\delta_{N+1} ,
\ee
and, for any $\alpha,\beta$ there exists a positive finite $M$ such that the following estimate holds (for simplicity we restrict to the case $\rho=0$) 
\be
\label{eq150}
\left|\partial_x^\alpha\partial_\xi^\beta \delta_{N+1}(x,\xi)\right|\leq C m(x,\xi)n(x,\xi) 
\\ 
\label{eq151}\times\left[ \sum_{j=0}^{N}\left(p_{mM}(a_j)p_{nM}(r_{N+1}(b))+p_{nM}(b_j)
p_{mM}(r_{N+1}(a)) \right)\right.
\\
\label{eq152}
\left. + \sum_{N\leq j+k\leq 2N} p_{mM}(a_j)p_{qM}(b_k)+p_{mM}(r_{N+1}(a))p_{nM}(r_{N+1}(b)) 
\right] 
\ee
where we denoted by $r_{N+1}(a)$ the remainder of the asymptotic expansion of $a$ truncated at order $N$ and we denoted
\be
\label{eq153}
p_{mM}(a)=\sup_{{|\alpha|+|\beta|\leq M, \ (x,\xi)}}\frac{|\partial^\alpha_x \partial_\xi^{\beta}a(x,\xi)|}{m(x,\xi)} . 
\ee
\end{theorem}

Fix a positive integer $s\geq 1$ denote $A=H_0^s$ (where $H_0$ is the operator \eqref{eq2}), from Theorem \ref {Theorem8} it follows that $A=Op^w(a)$ with a suitable $a$ having principal symbol $a_0=(\xi^2+V)^s$. \ Denote also $b:=\xi^{2s}+V^{s}$ and $B=Op^w(b)$. \ Since $V(x) \ge 1$ then one has
\be
\label{eq154}
\frac 1C \leq \frac{1}{C}{b(x,\xi)}\leq a_0(x,\xi)\leq Cb(x,\xi),\ \forall (x,\xi)\in\re^{2d} .
\ee

\begin{lemma}
\label{Lemma6} One has
\be
\label{eq155}
\norma{A\psi}_{L^2}\leq C\norma{B\psi}_{L^2},\quad \forall \psi\in D(B),
\\
\label{eq156}
\norma{B\psi}_{L^2}\leq C\norma{A\psi}_{L^2},\quad \forall \psi\in D(A).
\ee
\end{lemma}

\proof Consider $b^{-1} $, then $b^{-1}\in\Sigma^{b^{-1}}_0$. \ Denote $\B:=Op^w(b^{-1})$. \ By Theorem \ref{Theorem8} one has 
\be
\label{eq157}
A\B=Op^w\left(\frac{a_0}{b}\right)+\hbar Op^w(\delta_1)
\\
\label{eq158}
B\B=\uno+\hbar Op^w(\delta_2)
\ee
with $\delta_{1,2}$ estimated by \eqref{eq150}-\eqref{eq152}. \ Since $\frac{a_0}{b}$ is bounded together with its derivatives, it follows that $\Delta_1:=Op^w(\delta_1)$ is bounded. \ Similarly $\Delta_2:=Op^w(\delta_2)$ is bounded.

Thus, using Neumann formula one gets that the operator $\uno+\hbar\Delta_2$ is invertible provided $\hbar$ is small enough. \ So, from \eqref{eq158} one has 
\be
\label{eq159}
(\uno+\hbar\Delta_2)^{-1}B\B=\uno\quad\iff\quad\B^{-1}= (\uno+\hbar\Delta_2)^{-1} B.
\ee
Finally one has 
\be
\label{eq160}
\norma{A\psi}_{L^2}&=& \norma{A\B\B^{-1}\psi}_{L^2} \leq \norma{A\B}_{{\mathcal L}(L^2,L^2)} \norma{\B^{-1}\psi}_{L^2}  \\ 
\nonumber &\leq & C\norma{(\uno+\hbar\Delta_2)^{-1} B\psi  }_{L^2} \leq C\norma{B\psi}_{L^2} .
\ee
\qed

\section{Technical Lemmas}
 \label{appendiceB}

We start by a Lemma which is needed for the estimate of the solution of the homological equation. \ In its statement we will denote by $\ell^2$ the real Hilbert space of the sequences $(z_j)_{j\geq 3}$ endowed by the scalar product 
\bee
\langle z;z'\rangle:=\sum_{j\geq 3}z_jz'_j = \sum_\gamma \sum_{j\geq 3, \ j \in J_\gamma}z_jz'_j
\eee

\begin{lemma}
\label{Lemma7}
Let $F:\ell^2\to\ell^2$ be a bounded linear operator, assume that the corresponding matrix elements $F_{jl}$ are different from zero only if $j\in J_\gamma$ and $l\in J_{\gamma'}$ with $\gamma\not=\gamma'$. \ Define a new linear operator $G$ with matrix 
\be
\label{eq161}
G_{jl}:=\frac{F_{jl}}{i(\lambda_l-\lambda_j)}
\ee
Then there exists a positive
$C$ such that the following estimate holds
\be
\label{eq162}
\norma{G}_{{\mathcal L} (\ell^2 , \ell^2 ) }\leq \frac{C}{\hbar^{3/2}}\norma{F}_{{\mathcal L} (\ell^2 , \ell^2 ) } ,
\ee
\end{lemma}

\proof First we recall that $|\lambda_j - \lambda_l |\ge
C^{-1}\hbar$ if $j \in J_\gamma$ and $l\in J_{\gamma'}$ for $\gamma
\not= \gamma'$, and that $\# J_\gamma \le C/\hbar$. \ Fix $l \in
J_{\gamma'}$. \ First remark that
\be
\label{eq163}
\sum_{j}|F_{jl}|^2\leq \norma{F}^2_{{\mathcal L} (\ell^2 , \ell^2 ) }
\ee
then, by Schwartz inequality
\bee 
\sum_{j}|G_{jl}|\leq \left(\sum_{j}|F_{jl}|^2\right)^{1/2}\left(\sum_{j\in J_\gamma , \gamma
\not= \gamma'}\frac{1}{| \lambda_l-\lambda_j |^2}\right)^{1/2} 
\eee
Fix $l\in J_{\gamma'}$ and estimate
\be
\label{eq164}
\sum_{j}\frac{1}{| \lambda_l-\lambda_j |^2}= \sum_{\gamma\not=\gamma'} \sum_{j\in J_\gamma} \frac{1}{| \lambda_l-\lambda_j |^2}
\\
=\left(\sum_{\gamma=\gamma'\pm 1}+\sum_{\gamma=1}^{\gamma'-2}+ \sum_{\gamma\geq \gamma'+2}\right)\left( \sum_{j\in J_\gamma} \frac{1}{| \lambda_l-\lambda_j |^2} \right)
\ee
but, due to the choice of the numbers $E_\gamma$ one has
\be
\label{eq166}
\left|\lambda_j-\lambda_l\right|\geq 
\left\{ 
\begin{matrix} 
C^{-1}\hbar & {\rm if} & \gamma=\gamma'\pm1 \\
E_{\gamma'}-E_\gamma & {\rm if} & \gamma\leq \gamma'-2 \\
E_{\gamma-1}-E_{\gamma'} & {\rm if} & \gamma\geq \gamma'+2
\end {matrix} \right.
\ee
thus \eqref{eq164} is estimated by
\be
\label{eq167}
\left(\sup_{\gamma}\# J_\gamma \right)\left( \frac{2C^2}{\hbar^{2}}+\sum_{\gamma=1}^{\gamma'-2}\frac{1} {(E_{\gamma'}-E_\gamma)^2} +\sum_{\gamma\geq\gamma'+2} \frac{1} {(E_{\gamma-1}-E_{\gamma'})^2} \right)
\ee
Since the sums in \eqref{eq167} are convergent due to our choice of the sequence $E_\gamma$ on has
\bee 
\sum_{j}\frac{1}{| \lambda_l-\lambda_j |^2} \leq\frac{C}{\hbar^{3}}
\eee
which gives
\be
\label{eq168}
\sum_{j}|G_{jl}|\leq \frac{C}{\hbar^{3/2}}\norma {F}_{{\mathcal L} (\ell^2 , \ell^2 ) }
\ee
>From this one has
\bee
\norma{Gz}^2_{\ell^2}=\sum_{j}\left| \sum_{l}G_{jl}z_l\right |^2\leq \sum_{j}\left(\sum_{l}\left|G_{jl}\right| \right) \left(\sum_{l}\left|G_{jl} \right||z_l|^2\right) \\
= \left(\sum_{l}\left|G_{jl}\right| \right)\sum_j\left(\sum_{l}\left|G_{jl} \right||z_l|^2\right) \leq \frac{C\norma F^2_{{\mathcal L} (\ell^2 , \ell^2 ) }}{\hbar^3}\norma z^2_{\ell^2}
\eee
\qed

We report now some Lemmas from \cite{Bam99} which are needed for the proof of Lemma \ref{Lemma5}. \ Here $\mu$ will be small parameter (not the small parameter used in the main part of the text) that in applications will be replaced by $\epsilon^r$ with some $r$.

\begin{lemma}
\label{Lemma8}
Let ${\G}:B_\rho\to\C$, $\rho>0$ be a function whose Hamiltonian vector field is analytic as map from $B_\rho\to\X^s_{\C}$; fix a positive $\delta<\rho$. \ Assume $\mu\normag{{\G}}\rho<\delta$ and consider the flow $\phi^t$ of the corresponding Hamiltonian vector field. \ Then, for
$|t|\leq\mu$, one has
\begin{equation}
\label{eq169}
|\phi^t-\uno|_{\rho-\delta}\leq\mu\normag{{\G}}{\rho}\ .
\end{equation}
\end{lemma}

\proof It is just an application of the equality
\bee
\xi(t)-\xi(0)=\int_0^t\frac{d\xi}{dt}(s)ds=\int_0^tX_{\G}(\xi(s))ds\ .
\eee
\qed

\begin{lemma}\label{Lemma9}
Consider ${\G}$ as above and let $\Ha$ be an analytic function with vector field analytic in $B_{\rho}$, fix $0<\delta<\rho$ assume $\mu\normag{{\G}}\rho\leq\delta/3$, then, for $|t|\leq\mu$, one has
\bee 
\normag{\Ha\circ \phi^t}{\rho-\delta}\leq \left(1+\frac{3}{\delta}\mu\normag{{\G}}{\rho}\right)
\normag{\Ha}\rho 
\eee
\end{lemma}
\proof First remark that, since $\phi^t$ is a canonical transformation one has 
\be
\label{eq170}
X_{\Ha\circ \phi^t}(\xi)= d\phi^{-t}(\phi^t(\xi))X_{\Ha}(\phi^t(\xi))
\ee
from which
\bee 
X_{\Ha\circ \phi^t}(\xi)=\left(d\phi^{-t}(\phi^t(\xi))-\uno \right)X_{\Ha}(\phi^t(\xi)) +X_{\Ha}(\phi^t(\xi)) .  
\eee
To estimate the first term fix $\delta_1:=\delta/3$; we have
\bee
\sup_{\lceil \xi \rceil \le \rho-3\delta_1}{\lceil d\phi^{-t}(\phi^t(\xi))-\uno \rceil }\leq \sup_{\lceil \xi \rceil \le \rho-2\delta_1}{\lceil d\phi^{-t}(\xi)-\uno \rceil } \\
\leq \frac1{\delta_1} \sup_{\lceil \xi \rceil \le \rho-\delta_1}{\lceil \phi^{-t}(\xi)-\xi \rceil }\leq \frac\mu{\delta_1}\normag{{\G}}{\rho}.
\eee

Going back to $\delta$ and adding the trivial estimate of the second term one has the thesis.\qed

\begin{lemma}
\label{Lemma10}
Let ${\G}$ and $\Ha$ be as above, fix $0<\delta<\rho$, and assume $\mu\normag{{\G}}{\rho} <\delta/3$, then, for $|t|\leq\mu$ one has
\bee
\normag{\Ha\circ \phi^t-\Ha}{\rho-\delta}\leq \frac5\delta\normag{\Ha}\rho\mu\normag 
{{\G}}{\rho}
\eee
\end{lemma}

\proof One has
\bee 
X_{\Ha\circ \phi^t}(\xi)-X_{\Ha}(\xi)=\left(d\phi^{-t}(\phi^t(\xi))-\uno \right)X_
{\Ha}(\phi ^t(\xi)) +\left[X_{\Ha}(\phi^t(\xi))-X_{\Ha}(\xi)\right]\ .
\eee
The norm of the square bracket is easily estimated using Lagrange Theorem and the Cauchy inequality in order to bound $dX_{\Ha}$.  \ The other term was already estimated in Lemma \ref{Lemma9}, so we have the thesis. \qed

Although $\Ha_0$ has unbounded vector field the vector field of $\Ha_0\circ \phi^\mu-\Ha_0$ is bounded, more precisely we have

\begin{lemma}
\label{Lemma11}
Let $\Ha$, $\Ha_0$, $\F$ and $\G$ as in \S \ref {esp}; that is $\G$ is the solution of the Homological equation \eqref{eq88}; denote by $\phi^t$ the flow of the corresponding Hamiltonian vector field and 
\bee 
\ell (\xi):=\Ha_0(\phi^\mu(\xi))-\Ha_0(\xi)-\mu\left\{\Ha_0,{\G} \right\} , 
\eee
then the vector field of $\ell$ is analytic and, for any $\delta<\rho$, satisfies
\bee 
\normag{\ell}{\rho-\delta}\leq \mu^2\frac {25}{\delta} \normag{{\G}}\rho\normag{ \Ha }\rho .  \eee
\end{lemma}

\proof One has
\bee 
\Ha_0(\phi^\mu(\xi))-\Ha_0(\xi)=\int_0^\mu\frac d{dt} \Ha_0(\phi^t(\xi))dt = -\int_{0}^\mu \F (\phi^t(\xi))dt ,
\eee
where we used the homological equation (\ref {eq88}) to calculate $\left\{\Ha_0,{\G}\right\}$. \ Then one has
\bee
\ell(\xi)=\int_0^\mu\left [ \F(\phi^t(\xi))-\F(\xi) \right ] dt ;
\eee
Using Lemma \ref{Lemma4} and Lemma \ref{Lemma10} one gets the thesis.\qed

 \end {document}